\documentclass[twocolumn]{scrartcl}
\usepackage[
  left=2cm,
  right=2cm,
  top=2.5cm,
  bottom=2cm,
]{geometry}
\usepackage{tikz}
\usepackage[utf8]{inputenc}
\usepackage{amsmath}
\usepackage{todonotes}
\usepackage[colorlinks=false]{hyperref}
\hypersetup{ hidelinks = true, }
\usepackage{amsfonts,amssymb,amsthm, bbm,braket}

\newcommand{\R}{\mathbb{R}}

\usepackage[ruled, linesnumbered, vlined]{algorithm2e}
\usepackage[titletoc,toc,title,page]{appendix} 
\usepackage{babel}
\usepackage{csquotes}
\usepackage{xcolor}
\usepackage{pgfplots}
\usepackage{graphicx}
\usepackage{subfig}

\newcommand{\conjug}[1]{{\fontdimen8\textfont3=0.25pt\mkern2mu\overline{\mkern-1mu #1\mkern-2mu}}\mkern2mu}

\newtheorem{definition}{Definition}

\newtheorem{theorem}{Theorem}
\newtheorem{lemma}{Lemma}

\newtheorem{assumption}{Assumption}
\title{Algorithm Unfolding for Block-sparse and MMV
Problems with Reduced Training Overhead}
\author{Jan Christian Hauffen, Peter Jung, Nicole M\"ucke}
\date{\small\textit{j.hauffen@tu-berlin.de, peter.jung@tu-berlin.de, nicole.muecke@tu-braunschweig.de}}

\graphicspath{{arxiv_pictures}}

\begin{document}

\maketitle

\begin{abstract}\textit{
    In this paper we consider algorithm unfolding for the Multiple Measurement Vector (MMV) problem in the case where only few training samples are available. Algorithm unfolding has been shown to empirically speed-up in a data-driven way the convergence of various classical iterative algorithms but for supervised learning it is important to achieve this with minimal training data.
    For this we consider learned block iterative shrinkage thresholding algorithm (LBISTA) under different training strategies. To approach almost data-free optimization at minimal training overhead the number of trainable parameters for algorithm unfolding has to be substantially reduced. 
    We therefore explicitly propose a reduced-size network architecture based on the Kronecker structure imposed by the MMV observation model and present the corresponding theory in this context.
    To ensure proper generalization, we then extend the analytic weight approach by Lui et al to LBISTA and the MMV setting. Rigorous theoretical guarantees and convergence results are stated for this case.
    We show that the network weights can be computed by solving an explicit equation at the reduced MMV dimensions which also admits a closed-form solution. 
    Towards more practical problems, we then consider convolutional observation models and show that the proposed architecture and the analytical weight computation can be further simplified and thus open new directions for convolutional neural networks.
    Finally, we evaluate the unfolded algorithms in numerical experiments and discuss connections to other sparse recovering algorithms.}
\end{abstract}

\section{Introduction}
This paper connects the multiple measurement vector (MMV) problem, block- or joint-sparsity and recent results of deep unfolding of the iterative shrinkage thresholding algorithm (ISTA) to reconstruct unknown joint-sparse vectors from given linear observations. Such vectors could be for example signals received at the different antennas in a wireless communication problem or, in a computational imaging setup, discrete images observed at different detectors or aggregation stages. Compressed sensing is a way to reconstruct compressive measurements from their underdetermined systems and first theoretical breakthroughs were achieved by Candès, Romberg, Tao and Donoho \cite{candes2006robust,donoho2006compressed} leading to an approach where fewer samples can be used, than stated within the Nyquist–Shannon sampling theorem \cite{shannon1949communication}. They were able to show that unknown vectors can be reconstructed using convex optimization if the linear mapping fulfilled certain assumptions \cite{candes2006robust, candes2005decoding}. These idea rely on minimizing $\ell_1$-norm to promote sparsity and the approach of basis pursuit \cite{rudelson2005geometric}.  These convex optimization problems could then be solved with iterative algorithms, in \cite{figueiredo2007gradient} gradient projection approaches are presented and in \cite{fornasier2008iterative} the idea of thresholding algorithms, which will be also discussed in this work. Although this is already a well researched field, in practice leading to high computational effort because of many iterations and large underlying systems and is thus not suitable for real world applications. Thus Karol Gregor and Yann LeCun proposed to use the iterative structure of these algorithms for a neural network and to train each iteration step \cite{gregor2010learning}, which is also referred to as deep unfolding and will be discussed in Section 2.  Convergence for deep unfolding of the iterative thresholding algorithm has also been studied by Chen et al. \cite{ConvUnfoldedISTA}. Chen et al. proposed in \cite{ALISTA} to exclude the weight matrix from the data-driven optimization approach and pre-compute this by data-free optimization and thus presented Analytical LISTA (ALISTA). In recent results Chen et al. could reduce the training procedure even more, by showing that only tuning three hyperparameters is sufficient, proposing HyperLISTA \cite{chen2021hyperparameter}. Creating large sets of training data is often difficult in practice and thus it is important to reduce the trainable parameters. Therefore, we'll extend the already stated concepts for the block-sparse setting and especially for the multiple measurement vector problem, developing suitable learned algorithms, with only a few trainable parameters and similar theoretical guarantees as ALISTA.
\subsection{Multiple Measurement Vector Problems}
Multiple measurement vector (MMV) problems occur in many applications, for example in tomography \cite{gorodnitsky1995neuromagnetic}, communication \cite{9686735 }, blurred image reconstruction or superresolution \cite{ahmadi2020photothermal}. In the following we derive the connection to block-sparsity.
In an MMV problem we assume that we derived $d\in\mathbb{N}$ measurements $y^l\in\R^m$, $l=1,\dots,d$ from $d$ sparse signal vectors $x^l\in\R^{n}$ sharing the same support $supp(x^l)=\lbrace i\, : | x_i^l| \neq0 \rbrace$, which is referred to as joint-sparsity \cite{ziniel2012efficient, MMVRecovery}. The MMV problem can then be presented as solving the following $d$ equations
\begin{align}
    y^l = Kx^l+\tilde{\epsilon}  \, \label{eq::MMVvectorized}
\end{align}
for $l=1,\dots,d$ and with $K\in\R^{m\times n}$. This can be rewritten in the following matrix equation form
\begin{align}
Y = KX+\tilde{E} , \label{eq::MMVMatrix}
\end{align}
where $X=(x^1, \dots, x^{d})\in\R^{n\times d}$, $Y=(y^1, \dots, y^{d})\in\R^{m\times d}$. With the vectorizing operator $\text{vec}(\cdot)$, stacking each column of an matrix on each other, 
we can cast \eqref{eq::MMVMatrix} into a block-sparse problem. We have
\begin{align*}
Y &= KX\\
\Leftrightarrow Y^T & = X^T K^T \\
\Leftrightarrow \text{vec}(Y^T ) & = (K\otimes I_{d})\text{vec}(X^T),
\end{align*}
where we used the well-known vectorization property of matrix equations, see for example \cite{schacke2004kronecker}. Here $\otimes$ is the Kronecker product. The vector $x=\text{vec}(X^T)$ is block-sparse with $n$ blocks of length $d$, if the signals $x^l$ are jointly sparse. With $D=(K\otimes I_{d})\in\R^{n_y\times n_x}$, where $n_y=m\cdot d$ and $n_x=n\cdot d$, we obtain the block-sparse setting considered in this work.
\subsection{Block Sparsity}
In the more general setting we want to reconstruct an unknown vector $x\in\R^{n_x}$ from a given matrix $D\in\R^{n_y\times n_x}$ and given $y\in\R^{n_y}$
\begin{align}
y = Dx+\epsilon \label{underlyingEq}\, ,
\end{align}
with $n_x=nd, n_y = md$ for some $n,m,d\in\mathbb{N}$. We assume that noise $\epsilon\in\R^{n_y}$ is added to $Dx$. In applications we often have this problem is ill-posed, i.e. $n_y<n_x$ or $D$ is not invertible. We assume that $x$ is the concatenation of $n$ "smaller" vectors of length $d$, called blocks, i.e. $x[i]\in\R^d$,
\begin{align}
x^T=[\underbrace{x_1 \dots x_d}_{=x\left[1\right]} \underbrace{x_{d+1} \dots x_{2d}}_{=x\left[2\right]} \dots \underbrace{x_{n_x-d+1} \dots x_{n_x}}_{=x\left[n\right]} ]^T\, .  \label{concatx}
\end{align}
Following the notation in \cite{Eldar}  we define
\begin{align*}
\|x\|_{2,0}=\sum_{i=1}^n I(\|x[i]\|_2>0)\, ,
\end{align*}
where $ I(\|x[i]\|_2>0)=1$ if $ \|x[i]\|_2>0$ and equal to zero otherwise and the $\ell_2$-norm is defined as $\|x\|^2_2=\sum_{i=1}^n |x_i|^2$. We call $x\in\R^{n_x}$ $s$-block-sparse if $\|x\|_{2,0}\leq s$. Similar to \eqref{concatx} we can construct the matrix $D$ in \eqref{underlyingEq} from $n$ "smaller" matrices $D[i]\in\R^{n_y\times d}$, $i=1,\dots,n$
\begin{align*}
D = \left(D[1]\, D[2] \,\dots \,D[n]\right)\, .
\end{align*}
Without loss of generality we can assume that these blocks are orthonormal, see \cite{Eldar}, i.e. $D[i]^TD[i]=I_d$, where $I_d$ is the $d\times d$ identity matrix. This assumption simplifies the presentation of several statements below. 
\begin{definition} The block-coherence of a matrix $D\in\R^{n_y\times n_x}$ is defined as 
\begin{align}
\mu_b(D) = \max_{i\neq j}\frac{1}{d}\|D[i]^TD[j]\|_2\, .
\end{align}
\end{definition}
Here we use $\|A\|_2 = \sqrt{\lambda_{max}(A^TA)}$, where $\lambda_{max}$ denotes the largest eigenvalue of matrix $A$. Note that the block coherence can also be introduced with a normalization factor $1/\left(\|D[i]\|_2\|D[j]\|_2\right)$, but since we assume orthonormal blocks we can neglect this. This reduces to the already known coherence $\mu$ for $d=1$
\begin{align*}
\mu(D)=\max_{i\neq j } |D_{:,i}^TD_{:,j}|\,,
\end{align*}
where $D_{:,i}$ denotes the $i$th column of $D$, see \cite{donoho2005stable}.  
In \cite{Eldar} it is shown that $0\leq\mu_b(D)\leq\mu(D)\leq 1$ and it is possible to derive recovery statements for small $\mu_b$ similar to $\mu$, for details see \cite{Eldar}. We can consider also the cross block coherence, which compares two matrices and will be important in the following.
\begin{definition} For $B,D\in\R^{n_y\times n_x}$ with $B[i]^TD[i]=I_{d}$ the cross block coherence is defined as
\begin{align}
\mu_b(B,D) = \max_{i\neq j}\frac{1}{d}\|B[i]^TD[j]\|_2\, .
\end{align}
\end{definition}
Similar to ordinary basis pursuit and LASSO \cite{chen2001atomic, kutyniok2014compressed, foucart2017mathematical, rudelson2005geometric} we use the following $\ell_{2,1}$-LASSO to solve \eqref{underlyingEq}
\begin{align}
\min_{x\in\R^{n_x}} \frac{1}{2}\|Dx-y\|^2_2+\alpha \|x\|_{2,1}\label{underlyingMinl21}
\end{align}
where $\|x\|_{2,1}=\sum_{i=1}^n \|x[i]\|_2$ is the $\ell_{2,1}$-norm of $x$, which will promote block-sparsity for the solution of \eqref{underlyingMinl21}. This convex programm can be solved by the fixed point iteration known as the block iterative shrinkage thresholding algroithm (Block-ISTA/BISTA)
\begin{align}
    x^{(k)}=\eta_{\alpha\gamma}\left(x^{(k-1)}-\gamma\left[D^T\left(Dx^{(k-1)} - y \right) \right] \right)\, ,\label{eq::BISTA}
\end{align}
where $\eta_\alpha$ is the block-soft-thresholding operator, given as
\begin{align}
    \eta_\alpha \left(x\right)[i] = \max\left\lbrace0, 1-\frac{\alpha}{\|x[i]\|_2}\right\rbrace x[i].
\end{align}
BISTA is an already well studied proximal gradient algorithm based on the functional in \eqref{underlyingMinl21}. It is known that the fixed point iteration in \eqref{eq::BISTA} converges for $\gamma\in (0, \frac{1}{L})$, where $L=\|D\|_2^2$ is the Lipschitz constant of the least squares term in \eqref{underlyingMinl21} w.r.t to $x$, to a solution of \eqref{underlyingMinl21}, if at least one solution exists, see for example \cite{byrne2008applied, bauschke2011convex, Beck}. On the other hand the choice of the regularization parameter $\alpha$ has to be done empirically and is very crucial for a "good" recovery. If $\alpha$ is set too large, this can lead to too much damping, possibly setting blocks to 0 that actually have a non-zero norm . If $\alpha$ is too small we get reverse effects. In practice this leads to problems, since computing the iterations require high computational effort. Deep unfolding is a way to tackle these problems, i.e. reduce the number of iterations by learning optimal regularization parameters and step-sizes. There are already classical concepts in increasing the convergence speed by using an additional step in updating the current iterate $x^{(k)}$, by using the previous $x^{(k-1)}$, resulting in Block Fast ISTA \cite{Beck, combettes2011proximal}. On the other hand the choice of optimal parameters is still solved empirically.
\section{Deep Unfolding and Learned BISTA}
Recently, the idea of deep unfolding has been developed, where the goal is to optimize these parameters of such an iterative algorithm, \cite{gregor2010learning,kim2020element, ALISTA}.  This in turn gives us an iterative algorithm with optimal chosen step-size $\gamma$ and regularization parameters, but we will see that we don't have to restrict our self only to those parameters. Recently Fu et al. proposed Ada-BlockLISTA by applying deep unfolding to block-sparse recovery \cite{fu2021deep}. They show an increase in the convergence speed with numerical examples, but don't cover theoretical studies.\\
In the following we are going to present the idea of deep unfolding, then we are going to derive Learned BISTA (LBISTA).
\subsection{Deep Unfolding}
We will now formalize the concept of deep unfolding for an arbitrary operator that depends on a certain set of parameters, before we apply this to the previously presented fixed point iteration. To this end we define an operator
\begin{align}
T(\cdot\,;\, \theta,\, y) :X \to X
\end{align}
which depends on a set of parameters $\theta\in \Theta$, like the stepsize of a gradient descent operator and an input $y$. For example for BISTA this would be $\theta = ( \alpha, \gamma )$ with 
\begin{align*}
T(x\,;\, \theta,\,y) = \eta_{\alpha\gamma}\left(x-\gamma\left[D^T\left(Dx  - y \right) \right] \right) \, .
\end{align*}
We assume that $\text{Fix}(T\left(\cdot\,;\, \theta,\,y\right))\neq \emptyset$ and that we have convergence for the fixed point iteration 
\begin{align}
T^{k} \left(x^{(0)}\,;\, \theta,\,y\right)=x^{(k)} \label{fixedpointiter}
\end{align}
for an arbitrary $x^{(0)}\in X$. Deep unfolding now interprets each iteration step as the layer of a neural network and uses the parameters $\theta\in\Theta$ as trainable variables. In more detail this means we look at the $K$th iteration of \eqref{fixedpointiter}, i.e. the composition
\begin{align*}
	\underbrace{\left( T \circ \dots \circ T \right)}_{K\, \text{times}} (x^{(0)}\,;\, \theta,\,y) = x^{(K)}
\end{align*}
for some $K\in\mathbb{N}$.
By unfolding this iterative scheme we define
\begin{align*}
	T_{\theta^{(k-1)}}\left( \cdot\, ;\, y\right):= T\left( \cdot\, ;\,\theta^{(k)},\, y\right)
\end{align*}
for $k=1,\dots,K$ and set $\theta^{(k)}$ as the set of trainable variables in this operator, so they can vary in each iteration step. For example in LBISTA we will get $\theta^{(k)}=( \alpha^{(k)}, \gamma^{(k)} )$ in the later called tied case. With this we get the following composition
\begin{align}
	\left( T_{\theta^{(K-1)}} \circ \dots \circ T_{\theta^{(0)}} \right)  \left(x^{(0)}\, ; y\right) = x^{(K)}\, . \label{unfold}
\end{align}
and define the operator
\begin{align}
    \mathcal{T}_{\Tilde{\theta}} := T_{\theta^{(K-1)}} \circ \dots \circ T_{\theta^{(0)}} ,
\end{align}
where we get the full parameter space $\Tilde{\Theta}=\Theta\times \dots \times\Theta$, with trainable variables $\Tilde{\theta}=\bigcup_{i=1}^K\theta^{(i)} $.  $\mathcal{T}_{\Tilde{\Theta}}$ will then be the neural network which will be trained with respect to $\Tilde{\theta}$. It seems thus that deep unfolding can be applied to any iterative algorithm and help us to estimate the best choice of parameters, but we will only present deep unfolding for BISTA and consider deep unfolding for arbitrary operators in future work.

\newcommand{\loss}{\ell}
\subsection{Learning}
In this section we give an overview for the training procedure used in this work. The general idea of supervised learning is to choose model parameters such that the predictions are close, in some sense, to the unknown target, i.e., in our case the unknown vector $x$ in \eqref{underlyingEq} generating the measurement $y$. Hence, we aim to minimize an expected loss over an unknown distribution $\mathcal{D}$:
\begin{align}
\min_\theta \left[ R(\theta) :=  \mathbb{E}_{x^*\text{\raisebox{-0.9ex}{\~{}}}\mathcal{D}}\left[\loss(\hat{x}-x^*)\right]\right] \label{learning}
\end{align}
where $\loss(\cdot)$ is a given loss function, $\hat{x}=\mathcal{T}_{\Tilde{\theta}}  \left(x^{(0)}\, ; y\right)$ is the output of the model and $x^*$ is the ground-truth. Here, we will use the squared $\ell_2$-loss $\loss(x)=\frac{1}{2}\|x\|_2^2$. The objective functional $R(\theta)=\mathbb{E}_{x^*\text{\raisebox{-0.9ex}{\~{}}}\mathcal{D}}\left[\loss(\hat{x}-x^*)\right]$ is also called risk of the model. 
Since the underlying distribution $\mathcal{D}$ is unknown we take a batch of $S$ independently drawn samples of input and output data $(x_j^*,y_j)$ for $j=1,\dots,n_{train}$ according to \eqref{underlyingEq} and minimize instead data-driven the empirical risk
\begin{align}
R_S(\theta) = \frac{1}{n_{train}} \sum_{j=1}^{n_{train}}\loss(\hat{x}_j-x^*_j)\, . \label{emprisk}
\end{align}
Proceeding in this way for all layer at once is sometimes referred as end-to-end learning. Because of the special structure of our deep unfolding models, and inspired by  \cite{musa2021plug}, we instead train the network layer-wise, by optimizing only $\theta_{case}^{(k-1)}$ for layer $k$ yielding the following training procedure: Let $k\in\lbrace1,\dots,K\rbrace$, 
\begin{align*}
    \min_{\theta_{case}^{(k-1)}} \mathbb{E}_{x^*\text{\raisebox{-0.9ex}{\~{}}}\mathcal{D}}\left[\loss(\hat{x}^{(k)}-x^*)\right] \, ,
\end{align*}
where $\hat{x}^{(k)}$ is the output of the $k$th layer. We realized this training as follows, we generate a validation set $\left(x_{i, \text{validation}}^*, y_{i, \text{validation}}\right)$, used to evaluate the model while training  and a training set $\left(x_{i, \text{train}}^*, y_{i, \text{train}}\right)$, $i=1,\dots, n_{train}$, used to calculate \eqref{emprisk}. This objective is locally minimized by gradient descent methods. As a stopping criteria we evaluate the normalized mean square error, defined as 
\begin{align*}
NMSE(x,\hat{x}^{(k)}) = \frac{\|\hat{x}^{(k)}-x^*\|_2^2 }{\|x^*\|_2^2} \, ,
\end{align*}
depending on the validation set, and stop if the maximum of all evaluated $NMSE$ stays the same for a given number of iterations. See Algorithm \ref{Training}, where $Adam$ is the ADAM Optimizer \cite{kingma2017adam} depending on an training rate $t_r$ and the functional which should be minimized with respect to given variables, here the loss function $\loss(\cdot)$ with respect to $\theta^{k-1}$.
\\ \hspace{2pt} \\
\begin{algorithm}
 \For{$k\leq K$}{
	\While{$\neg ( NMSE(x^{(k)}, x_{i, \text{validation}}^*)$ < $tol$ for 5000 Iterations)}{
	
	$Adam\left(t_r\,, l(\hat{x}^{(k)}-x_{i, \text{train}}^*)\,, \theta^{(k-1)}\right)$\\
	with $x^{(k)}=\left( T_{\theta^{k-1}} \circ \dots \circ T_{\theta^0} \right) (x_0; \, y_{i, \text{train}})$ for all $i=1,\dots, n_{train}$
	}
 }
 \caption{Training}\label{Training}
\end{algorithm} \hspace{2pt} \\
\subsection{Learned BISTA}
In the following we present four different unfolding techniques for BISTA. We present a {\em tied} (weights are shared between different layers) and {\em untied} (individual weights per layer) case, which refers to different training approaches for the matrices
\begin{align*}
    S & = I - B^TD \\ 
    B & = \gamma D.
\end{align*}
\textbf{Tied LBISTA: }
The idea of LBISTA is now to fix the matrices $S$ and $B$ for all layers, but also include them in our set of trainable variables: 
\begin{align}
    x^{(k)}=\eta_{\alpha^{(k-1)}}\left(Sx^{(k-1)}+B^Ty\right) \, .\label{tiedLBISTA}
\end{align}
For LBISTA \eqref{tiedLBISTA} we get trainable variables \begin{align*}
    \theta = \left( \left(\alpha^{(k)}\right)_{k=0}^{K-1}, S,B \right).
\end{align*}
Algorithm \eqref{tiedLBISTA} is also referred to as vanilla LISTA in the sparse case.
Inspired by the LISTA-CP model, i.e. LISTA with coupled parameters, proposed in \cite{ALISTA} we will also consider LBISTA-CP 
\begin{align}
x^{(k)}=\eta_{\alpha^{(k-1)}}\left(x^{(k-1)}-\gamma^{(k-1)} B^T\left(Dx^{(k-1)}-y\right) \right) .\label{TiLBISTA}
\end{align}
For LBISTA-CP \eqref{TiLBISTA} we get 
\begin{align*}
    \theta =  \left( \left(\alpha^{(k)}\right)_{k=0}^{K-1}, \left(\gamma^{(k)}\right)_{k=0}^{K-1}, B \right).
\end{align*}
\textbf{Untied LBISTA: }
The idea of untied LBISTA is then to use in each layer different matrices $S$ and $B$ to train, i.e., \\ \begin{align}
    x^{(k)}=\eta_{\alpha^{(k-1)}}\left(S^{(k-1)}x^{(k-1)}+\left(B^{(k-1)}\right)^T y  \right) .\label{untiedLBISTA}
\end{align}
For LBISTA (untied) \eqref{untiedLBISTA} we get trainable variables
\begin{align*}
    \theta = \left( \left(\alpha^{(k)}\right)_{k=0}^{K-1},  \left(S^{(k)}\right)_{k=0}^{K-1}, \left(B^{(k)}\right)_{k=0}^{K-1} \right).
\end{align*}
Inspired by algorithm \eqref{TiLBISTA} we will also consider algorithm \eqref{LBISTA-CPSS}, which will be referred to as LBISTA-CP (untied), \cite{ConvUnfoldedISTA, ALISTA}:
\begin{align}
    \nonumber&x^{(k)}=\\&\eta_{\alpha^{(k-1)}}\left(x^{(k-1)}-\gamma^{(k-1)}\left(B^{(k-1)}\right)^T\left(Dx^{(k-1)}-y\right)\right). \label{LBISTA-CPSS}
\end{align}
For LBISTA-CP (untied) \eqref{LBISTA-CPSS} we get 
\begin{align*}
    \theta =  \left( \left(\alpha^{(k)}\right)_{k=0}^{K-1}, \left(B^{(k)}\right)_{k=0}^{K-1}  \right).
\end{align*}
Hence, compared to $\mathcal{O}(n_yn_x + K)$ parameters in the tied case, now more training data and longer training time is required to train now $\mathcal{O}(Kn_yn_x+K)$ parameters.

We initialize the trainable variables with values from original BISTA. In \cite{ConvUnfoldedISTA} it has been shown that convergence of LISTA-CP (untied) can be guaranteed if the matrices $B^{(k)}$ belong to a certain set and their proof can be extended to block-sparsity. The steps are very similar to the convergence proof for Learned Block Analytical ISTA given in the next section. 

\section{Analytical LBISTA}
In the previous section we presented several approaches for learned BISTA where optimal weights are optimized in a data-driven fashion. In \cite{ALISTA} Liu et al. instead proposed to analytically pre-compute the weights and only train step-size and threshold parameters. It turns out that this {\em Analytic LISTA} (ALISTA) with the so called analytical weight matrix is as good as the learned weights. In the following we are going to extend and improving the theoretical statements for ALISTA to the block-sparse case and propose Analytical LBISTA. In contrast to \cite{ALISTA}, we will provide a direct solution and also show different ways to calculate the analytical weight matrix in different settings.

\subsection{Upper and Lower Bound}
This part of the paper will focus on combining and extending several theoretical statements from \cite{ConvUnfoldedISTA, ALISTA} and applying these for the block-sparse case. With the following two theorems we are then going to present Analytical LBISTA, by showing that this is as good as LBISTA-CP (untied) with a pre-computed $B$.
\subsubsection{Upper Bound}
In this section we start with an upper bound for the error of the approximation generated by \eqref{LBISTA-CPSS},
i.e. LBISTA-CP (untied), and the exact solution $x^*$ for given parameters.  For this we modify Assumption 1 from \cite{ALISTA} to be consistent with the block-sparse setting.
\begin{assumption}We assume $(x,\epsilon)\in\mathcal{X}_b(M,s,\sigma)$ with:
  \begin{align}   
    (x,\epsilon)
    &\in\mathcal{X}_b(M,s,\sigma)\\
    & = \lbrace x :\|x[i]\|_2\leq M \, \forall i=1,\dots,n, \nonumber\\
    &\quad \|x\|_{2,0} \leq s\, , \|\epsilon\|_2\leq \sigma \rbrace.\nonumber
  \end{align}
\end{assumption}
As already mentioned, a matrix with small block-coherence has good recovery conditions. In \cite{ALISTA} Liu et al. propose Analytical LISTA, where the pre-computed matrix $B$ is minimizing the mutual cross-coherence. This motivates the following definition.
\begin{definition} With $D\in\R^{n_y \times n_x}$ we define the generalized mutual block-coherence
\begin{align}
\tilde{\mu}_b(D) & = \underset{\underset{\underset{\forall 1\leq l \leq n}{B^T[l]D[l]=I_d}}{B\in\R^{n_y\times n_x}}}{inf}\left\lbrace \underset{\underset{1\leq i,j \leq n}{i \neq j }}{max}
\frac{1}{d} \| B^T[i]D[j]\|_2\right\rbrace \nonumber\, \\
& = \underset{\underset{\underset{\forall 1\leq l \leq n}{B^T[l]D[l]=I_d}}{B\in\R^{n_y\times n_x}}}{inf}\left\lbrace\mu_b\left(B,D\right)\right\rbrace \, .\label{mutblockcoh}
\end{align}
We define, analogously to \cite{ALISTA} with $\mathcal{W}_b(D)$ the set of all $B\in\R^{n_y\times n_x}$ which attain the infimum in (\ref{mutblockcoh}), i.e. $\mathcal{W}_b(D)=\left\lbrace B\in\R^{n_y\times n_x}\, :\, \mu_b(B,D)=\tilde{\mu}_b(D) \right\rbrace$.
\end{definition}
Note that the set $\mathcal{W}_b(D)$ is non empty, because the set of feasible matrices $\lbrace B\in\R^{n_y\times n_x} : B^T[i]D[i]=I_d ,  1\leq i \leq n\rbrace$ contains at least $D$ because we assume $D^T[i]D[i]=I_d$, also $0\leq \tilde{\mu}_b(D) \leq \mu_b(D)$ and therefore \eqref{mutblockcoh} is a feasible and bounded program, see supplementary material to \cite{ConvUnfoldedISTA}. We will call matrices from $\mathcal{W}_b(D)$ analytical weight matrices.
\begin{definition} The block-support of a block-sparse-vector $x\in\mathcal{X}_B(b,s)$ is defined as
\begin{align}\label{eq::blocksupp}
\text{supp}_{b}\left(x\right)=\left\lbrace i : \|x[i]\|_2\neq 0\,  \forall i=1,\dots n \right\rbrace \, .
\end{align} 
\end{definition}
We will now derive an upper bound for the $\ell_2$-error and thus showing convergence of LBISTA-CP for a special matrix $B$ and given parameters $\alpha^{(k)}$ and $\gamma^{(k)}$.  In \cite{ConvUnfoldedISTA} Liu et al. showed linear convergence for unfolded ISTA with additional noise, more precivesly for LISTA-CP (untied), if the matrices $B^{(k)}$ belong to a certain set. In \cite{ALISTA} it was shown that we can pre-compute such a matrix $B$, chose $B^{(k)}=B$, chosen by a  data-free optimization problem and still have the same performance. For this new proposed unfolded algorithm linear convergence was also shown, without additional noise. In \cite{ALISTA} convergence only in the noiseless case was shown, but the results derived in \cite{ConvUnfoldedISTA} were derived with bounded noise. Thus we are going to combine these two proofs and extend it to block-sparsity:

For given $(x,\epsilon)$, $y=Dx+\epsilon$ and parameters $\{\theta^{(k)}\}_{k=1}^K$, we abbreviate with $\lbrace x^{(k)}(x,\epsilon) \rbrace_{k=1}^K $ the sequence generated by \eqref{LBISTA-CPSS} with $x^{(0)}=0$. Further, we define 
\begin{equation}
\begin{split}
    C^{(k)}_{\mathcal{X}} 
        &= \sup_{(x,\epsilon)\in\mathcal{X}_b(M,s,\sigma)} \lbrace \|x^{(k)}(x,\epsilon)-x\|_{2,1} \rbrace\\ 
    C & =\underset{k}{\sup} \underset{j = 1,..., n}{\max} |\gamma^{(k)}|\|B[j] \|_2
\end{split}
\end{equation}
\begin{theorem}
\label{UpperBound}
For any $B\in\mathcal{W}_b(D)$ and any sequence $\gamma^{(k)}\in \left(0, \frac{2}{\mu(2 s-1)+1}\right)$ and parameters $\left( B^{(k)}:=B, \alpha^{(k)},\gamma^{(k)}\right)$, for $k\leq K$, with:
\begin{equation}
    \frac{\alpha^{(k)}-C\sigma}{\gamma^{(k)}\mu(D)C^{(k)}_{\mathcal{X}}}\in[1,\kappa]
    \label{eq:thm1:alphak}
\end{equation} 
for some $\kappa\geq1$, with $\mu=d\tilde{\mu}_b(D)$.
 With $M>0$ and $s<(\mu^{-1}+1)/2$ we have:
\begin{align}
\text{supp}_{b}&\left(x^{(k)}(x^*)\right) \subset \text{supp}_{b}\left(x^*\right) =: \mathbb{S}\, , \label{support2} \\
\|&x^{(k)}-x^* \|_2 \leq  \exp \left(-\sum_{\tau=0}^{k-1} \tilde{a}(\tau)\right) sM\label{upperbound2}\\&+C\sigma\left(1+\sum_{\tau = 0}^{k-1} \exp\left(-\sum\limits_{s = \tau}^{k-\tau}\tilde{a}(s)\right) \right) \, .\nonumber
\end{align}
where 
\begin{align*}
    \tilde{a}(\tau)&=-\log\left(\gamma^{(\tau)}\mu\left((\kappa+1) s-1\right)+|1 - \gamma^{(\tau)}| \right)\\&>0\, .
\end{align*}
\end{theorem}
Note that with $\kappa = 1$ we obtain the results in \cite{ALISTA} but one can't always meet this condition. But we need at least $\alpha^{(k)} \geq d\gamma^{(k)}\tilde{\mu}_b(D)C^{k()}_{\mathcal{X}}+ C \sigma$ to have \eqref{support2}. On the other hand, one can always find such a $\kappa$ from the trained parameters and thus use therefore the theorem afterwards. Obviously a worse $\alpha$ effects the upper bound of the $\ell_2$-error and thus appears in $\tilde{a}(\tau)$. Summarizing, above theorem shows now convergence on the training set, even if $\kappa \neq 1$.

\subsubsection{Lower Bound}
This section states the lower bound for the $\ell_{2,1}$-error, showing that for convergence in the $\ell_{2,1}$-norm the defined parameters in Theorem \ref{UpperBound} are optimal chosen.
We now modify Assumption 2 from \cite{ALISTA} to be consistent with the block-sparse setting.
\begin{assumption}
$x^*$ is sampled from $P_X$. $P_X$ satisfies: $2\leq \mathbb{S} \leq s$ and $\mathbb{S}$ is uniform distributed over the whole index set. The non-zero blocks of $x^*$ satisfy the uniform distribution and $\|x[i]\|_2\leq M$ for all $i\in\mathbb{S}$. And we assume $\epsilon = 0$.
\end{assumption}
The latter Theorem states that the analytical weight matrix should minimize the generalized mutual block coherence. Therefore, for a lower bound, we will only consider matrices that are bounded away from the identity.
\begin{definition}
With $D\in\R^{n_y\times n_x}$, $s\leq 2$, $\Bar{\sigma}_{\min}>0$ we set 
\begin{align}
    \Bar{\mathcal{W}}&\left(D,s,\Bar{\sigma}_\text{min} \right) := \lbrace B\in\R^{n_y\times n_x} \\ &: \sigma_{\text{min}} \left(I-\left(B[(j\in\mathbb{S})]\right)^T D[(j\in\mathbb{S})]\right)\geq \Bar{\sigma}_{\min} \rbrace\, .\nonumber \label{eq::Wset}
\end{align}
\end{definition}
The parameters are chosen from the following set.
\begin{definition}\label{Def::nofalse}
Let $\lbrace x^{(k)}\rbrace_{k=1}^\infty$ be  generated by $x^{(k+1)} = \eta_{\alpha^{(k)}}\left(x^{(k)}-(B^{(k)})^T(Dx^{(k)}-y) \right)$ with parameters $\lbrace B^{(k)}, \alpha^{(k)}\rbrace_{k=0}^\infty$ and $x^{(0)}=0$. We define the set of all parameters guaranteeing no false positive blocks in $x^{(k)}$ by
\begin{align}
    \mathcal{T}=&\lbrace\lbrace B^{(k)}\in\Bar{\mathcal{W}}\left(D,s,\Bar{\sigma}_\text{min}\right), \theta^{(k)}\rbrace_{k=0}^\infty\,:\\ &\text{supp}_{B}\left(x\right)\subset \mathbb{S}, \, \forall x^* \in \mathcal{X}_b(M,s,0)\, , \forall k\rbrace\, .\nonumber
\end{align}
\end{definition}
This set is non-empty, since \eqref{support2} holds true if $\alpha^{(k)}$ are chosen large enough. Following mainly the proof in \cite{ALISTA} by extending the setting from sparsity to block-sparsity, the lower bound for the $\ell_{2,1}$-norm can be stated as follows. 
\begin{theorem}\label{Thm2:LowerBound}
Let $\lbrace x^{(k)}\rbrace_{k=1}^\infty$ be  generated by $x^{(k+1)} = \eta_{\alpha^{(k)}}\left(x^{(k)}-(B^{(k)})^T(Dx^{(k)}-y) \right)$. Under Assumptions 2, $\lbrace W^{(k)}\in\Bar{\mathcal{W}}\left(D,s,\Bar{\sigma}_\text{min}\right), \theta^{(k)}\rbrace_{k=0}^\infty\in\mathcal{T}$ and $\epsilon>0$, we have
\begin{align}
    \|x^{(k)}-x^*\|_{2,1}\geq \epsilon \|x^*\|_2\exp(-ck) \, ,\label{LowerBoundWithoutNoise}
\end{align}
with probability $1-\epsilon s^\frac{3}{2}-\epsilon^2$ and $c=\log 3-\log\bar{\sigma}_{min}$.
\end{theorem}
\subsection{Analytical LBISTA}
Analogously to \cite{ALISTA} and following the previous two theorems, decompose LBISTA-CP (untied), Algorithm \ref{TiLBISTA}, into two steps:
\begin{align}
x^{(k)}=\eta_{\alpha^{(k-1)}}\left(x^{(k-1)}-\gamma^{(k-1)}\tilde{B}^T\left(Dx^{(k-1)}-y\right) \right) \, , \label{eq::ALBISTA}
\end{align}
where, in the first step, $\tilde{B}$ is pre-computed, such that 
\begin{align*}
 \mu_b(\tilde{B},D)=\tilde{\mu}_b(D)\, .
\end{align*}
In the second step the parameters $\theta = \left(\left(\alpha^{(k)}\right)_{k=0}^{K-1}, \left(\gamma^{(k)}\right)_{k=0}^{K-1} \right)$ are trained layer wise, as discussed in the previous section. This results in a comparable method, with only $\mathcal{O}(K)$ trainable parameters, instead of $\mathcal{O}(n_y n_x+K)$ for LBISTA-CP \ref{TiLBISTA} or even $\mathcal{O}(Kn_y n_x+K)$ for LBISTA (untied) \ref{untiedLBISTA}.
\section{Computing the Analytical Weight Matrix}
ALBISTA relies on the analytical weight matrix, deriving this matrix can be challenging in practice, thus this section focuses on computing this matrix. 
We follow the procedure in \cite{ALISTA} by estimating \eqref{mutblockcoh} with an upper bound. But in addition to \cite{ALISTA} we state a closed form for the upper bound, currently this is done by a projected gradient descent approach. 
\subsection{Solving An Upper Bound}
Since the objective in \eqref{mutblockcoh} is not differentiable one solves the following upper bound problem
\begin{align}
     &\min_{B\in\R^{n_y\times n_x}}\frac{1}{d}\|B^T D\|_F^2 \label{reducedmutblockcoh} \\&\text{ s.t. } B^T[i]D[i]=I_d \text{ for } i = 1,\dots,n\, .\nonumber
\end{align}
This is derived from the following inequality
\begin{align*}
    \max_{i\neq j} \frac{1}{d}\|B^T[i]D[j]\|_2^2 &\leq \max_{i\neq j} \frac{1}{d}\|B^T[i]D[j]\|_F^2\\& \leq \frac{1}{d}\sum_{i,j} \|B^T[i]D[j]\|_F^2 \\
    & = \frac{1}{d} \|B^TD\|_F^2\,.
\end{align*}
In \cite{ALISTA} this is solved by a projected gradient method, but the following Theorem states a closed form of the solution of \eqref{reducedmutblockcoh}.
\begin{theorem}\label{thm::AnalyticalMatrix-Up}
The minimizer $B\in\R^{n_y\times n_x}$ of \eqref{reducedmutblockcoh} is given as the concatenation
\begin{align*}
B = \left(B[1]\, , B[2] \, , \dots,\, B[n]\right)\, ,
\end{align*}
where the $n$ blocks are given as
\begin{align*}
 B[i] = K_i^+\left(D[i]-E_i H_i \right)
\end{align*}
with 
\begin{align*}
	K_i & = \left(2DD^T\right)^2+D[i]D[i]^T \, , \\
	E_i & = 2DD^TD[i] \, , \\
	R_i & = D[i]-2DD^TK_i^+E_i \, , \\
	S_i & = - D[i]^TK_i^+E_i \, , \\
	L_i & = R_i^TR_i+S_i^TS_i \, , \\
	M_i & = K_i^+E_i(I-L_i^+L_i) \, , \\
	H_i & = L_i^+S_i^T+(I-L_i^+L_i)(I+M_i^TM_i)^{-1}\\&(K_i^+E_i)^TK_i^+(D[i]-E_iL_i^+S_i^T)\, , 
\end{align*}
for $i = 1,\dots, n$.
\end{theorem}
The proof can be found in Appendix \ref{proof:upperbound}. \\ 
Let $d=1$ and the singular value decomposition of $D$ given as $D = V \Sigma U^T$ and assume $B = V \tilde{\Sigma} U^T$. Then the solution of \eqref{reducedmutblockcoh} is given in an even simpler form, since
\begin{align*}
    \|B^TD\|_F^2& = \| U \tilde{\Sigma} V^T V \Sigma U^T\|_F^2 \\
    & = \| U \tilde{\Sigma}\Sigma U^T\|_F^2\\& = \| \Sigma^+\Sigma \|_F^2.
\end{align*}
Choosing $B = D^{+,T} diag(\tilde{d})^{-1}$, where $\tilde{d}=diag(B^{+}D)$, yields also a solution for \eqref{reducedmutblockcoh}. Here $diag$ follows the matlab/python notation, where $diag$ of a matrix gives the vector of the main diagonal and gives a diagonal matrix with a given vector on its main diagonal. 
For $d\geq 2$ the orthonormal block constraints would not be met, thus not yielding a feasible solution,.
\subsection{Computing the Analytical Weight Matrix in a MMV Problem}
In practice, often large data sets are obtained, i.e. by a large amount of measurements or  measurements $y^l, x^l$ representing pictures. For instance $1000$ pixels and $200$ measurements lead to a matrix $D$ with $(1000\cdot 200)^2$ elements.  Applying the theory for Analytical LBISTA could thus be difficult in practice. Although D is sparse, it can take a long time to calculate the analytical weight matrix. The following Theorem states the connection between the MMV setting and the block sparse setting for ALBISTA, showing that it's sufficient to minimize the generalized mutual coherence \eqref{eq::gen_coh}, see \cite{ALISTA}, for $K$ instead of minimizing the generalized mutual block-coherence \eqref{mutblockcoh} for $D=K\otimes I_d$.
\begin{theorem}\label{thm:mmvTrick1}Let $\tilde{B}$ attain the minimum in
\begin{align}
    \underset{\underset{\underset{\forall 1\leq i \leq n}{\tilde{B}_{:,i}^TK_{:,i}=1}}{\tilde{B}\in\R^{n\times n}}}{inf}    \underset{i\neq j}{\text{max}}\, |\tilde{B}_{:,i}^TK_{:,j}|\, , \label{eq::gen_coh}
\end{align}
where $\tilde{B}_{:,i}$ refers to the $i$th column. Then the minimum of
\begin{align*}
     \underset{\underset{\underset{\forall 1\leq l \leq n}{B^T[l]D[l]=1}}{B\in\R^{n_y\times n_x}}}{inf}\left\lbrace \underset{\underset{1\leq i,j \leq n}{i \neq j }}{max}
\frac{1}{d} \| B^T[i]D[j]\|_2\right\rbrace \, ,
\end{align*}
is given as $B=\tilde{B}\otimes I_{d}$, if $D=K\otimes I_d$.

\end{theorem}
The proof can be found in Appendix \ref{proof:mmvTrick}. Moreover the following relation holds, if $D=K\otimes I_d$,
\begin{align*}
    \mu_b(D) = \frac{1}{d}\mu(K),
\end{align*}
thus the block-coherence of $D$ can be enhanced by increasing the number of measurements $d$. Also it is feasible to solve \eqref{reducedmutblockcoh} by  the pseudo inverse in the MMV setting, since this is solved for $K$, i.e. $d=1$ 
\subsection{Circular Matrix Case}
Consider now the following setting, where the measurements $y^l\in\mathbb{R}^n$ are obtained by a circular convolution of $x^l$ with vector $k$, i.e.
\begin{align}
    y^l &= k\circledast x^l\label{eq::CircConv} \\ &= Kx^l, \, l=1,\dots, d \, , \nonumber
\end{align}
where $K$ is a circular matrix generated by a vector $k\in\R^n$, i.e. $K=circ(k)\in\R^{n\times n}$. Applying Theorem \ref{thm::AnalyticalMatrix-Up} to the circular case yields the following lemma.
\begin{lemma}
Let $k\in\R^{n}$ and let $B=circ(b)\in\R^{n\times n}$ where $b\in\R^n$ is given by 
\begin{align}
    \left(\begin{array}{cc}
        2KK^T & k  \\
        k^T & 0
    \end{array}\right)
    \left(\begin{array}{cc}
        b  \\
        \lambda 
    \end{array}\right) = \left(\begin{array}{cc}
        0 \\
        1
    \end{array}\right) \, , \label{eq::SolUpperBoundCirc}
\end{align}
$\lambda\in\R$. Then $B$ attains the minimum in \eqref{reducedmutblockcoh}.
\end{lemma}
The latter statement implies a simpler way to compute $b\in\R^n$ by using singular value decomposition 
\begin{align*}
    B^*K&=U^*diag\left(\sigma (B)\right)UU^*diag\left(\sigma (K)\right)U\\
    & = U^*diag\left(\sigma (B)\odot\sigma (K)\right)U \, ,
\end{align*}
where $\sigma(B)\in\mathbb{R}^n$ is the vector of singular values of $B$, respectively $K$, filled with zeros and $U$ an unitary matrix.
With  $U=1/\sqrt(n)F$, where $F$ is the Fast Fourier Transform (FFT) matrix, this leads to the conclusion $\sigma(B)=Fb = \hat{b}$, and thus 
\begin{align}
    b = F^{-1}\left(\conjug{1/\hat{k}}\right).
\end{align}
The expression $1/\hat{k}$ should be interpreted point wise and to be zero if $\hat{k}_i=0$. This also concludes that the computation of $B$ tends to be difficult in practice if $K$ has not full rank, since this means that $\hat{k}$ has at least one zero entry. In this case b has to be scaled with $\frac{n}{rank(K)}$, since
\begin{align*}
    b^Tk & = \frac{1}{n}\hat{b}^T\hat{k} 
         = \frac{1}{n}\|\hat{k}\|_0 \\
         & = \frac{rank(K)}{n} \overset{!}{=}1.
\end{align*}
\subsection{Toeplitz Matrix Case}
Considering the more general convolutional setting,
\begin{align}
    y = k\ast x \, ,\label{eq:genConv}
\end{align}
where $k\in\mathbb{R}^{\tilde{m}}$ and $x\in\mathbb{R}^n$ with $\tilde{m}<n$. This results in a Toeplitz matrix $K$
\begin{align*}
    K = \left(\begin{array}{cccc}
        k_1 & 0 & \hdots & 0  \\
        k_2 & k_1 & & 0 \\
        \vdots & \vdots & \ddots & \vdots \\
        k_{\tilde{m}} & k_{\tilde{m}} & & \vdots \\
        0 & k_{\tilde{m}} & & \vdots \\
        \vdots & \vdots & \ddots & \vdots \\
        0 & 0&\hdots  &k_{\tilde{m}}
    \end{array}\right)\in\mathbb{R}^{m\times n}.
\end{align*}
The reasoning of the previous section cannot be applied to show that the solution of \eqref{reducedmutblockcoh} must be a Toeplitz matrix. But the following can be observed: Let $b$ be constructed as discussed for $\tilde{K} = circ(\underline{k})$, where $\underline{k}$ is the concatenation of $k$ and a zero vector of suitable dimensions, i.e. the first column of $K$. The analytical weight matrix $B$ w.r.t. $K$ can be constructed as 
\begin{align*}
    B_{:,i} = T^{i-1}b, \, i=1,\dots, n\, ,
\end{align*}
where $T$ is the $m\times m$ cyclic shift matrix, i.e. only the $m\times n $ submatrix of $\tilde{B}= circ(b)\in\mathbb{R}^{m\times m }$ is used. The columns of $K$ can also be expressed through the cyclic shift of $\underline{k}$. Hence
\begin{align*}
    B_{:,i}^TK_{:,i} = b^T T^{-(i-1)}T^{i-1}k = b^Tk = 1,
\end{align*}
i.e. $B$ is a feasible solution of \eqref{reducedmutblockcoh} for $K$.
On the other hand the cross coherence is bounded, since
\begin{align*}
    \max_{i,j\leq n.\, i\neq j}|B_{:,i}^TK_{:,j}| & = \max_{i,j\leq n,\, i\neq j} |b^T T^{-(i-1)}T^{j-1}k | \\
    & =\max_{i,j\leq n,\, i\neq j} | b^T T^{j-i} k | \\
    & \leq \max_{i,j\leq m,\, i\neq j} | b^T T^{j-i} k |\\
    & = \max_{i,j\leq m,\, i\neq j}|\tilde{B}_{:,i}^T\tilde{K}_{:,j}|\\
    & = \|\tilde{B}^T\tilde{K} - I_m\|_\infty = 0 .
\end{align*}
Note: To have this upper bound $\tilde{K}$ needs to have full rank, which is the case if the full time continuous FT of $k$ has no zero points. Or one has to adjust the discrete grid. Thus constructing $B$ by extending $K$ to a circular matrix is a feasible approach.
\subsection{Connection to CNNs}
It is known that using the FFT in CNNs can increase the computation time, if the convolutional filter is big \cite{pratt2017fcnn, chitsaz2020acceleration}. In \cite{pratt2017fcnn} Pratt et al. showed that training weights in the Fourier domain can reduce training time while maintaining efficiency. By using Fourier Convolution the costs of $\mathcal{O}(n^2)$ operations could be reduced to $\mathcal{O}(n\log(n))$ operations. To connect the theory of FFT-CNNs and of unrolling ISTA in the context of deconvolution \eqref{eq::CircConv} or \eqref{eq:genConv} the gradient step can be viewed as follows
\begin{align}
    x & - \gamma b\ast\left(k\ast x - y\right)\nonumber \\
    & = x - \gamma b\ast k\ast x + \gamma b\ast y \nonumber \\
    & = \left(e - \gamma b\ast k\right)\ast x + \gamma b\ast y\nonumber \\
    & = f(\gamma)\ast x+\gamma \Tilde{b}\, .
\end{align}
This can be interpreted as a convolutional layer with kernel $f(\gamma)=\left(e - \gamma b\ast k\right)$ and bias $\Tilde{b}=b\ast y$, where $e=[1, 0,  \dots, 0]$. This means that ALISTA, with a Toeplitz matrix, can be interpreted as a CNN only two trainable parameters per layer, $\gamma, \lambda$. In the setting of FFT-CNN the update rule can be formulated as
\begin{align*}
    x & - \gamma b\ast\left(k\ast x - y\right) \\ 
     & = F^{-1}\left((\hat{e}-\gamma\hat{b}\odot\hat{k})\odot \hat{x} \right) + \gamma\Tilde{b} \, .
\end{align*}
On the other hand using FFT CNNs shows only a speed up if we deal with large data sets, i.e. by evaluating high resolution images, or  large filters, i.e. if $\tilde{m}$ is greater than $\log(n)$.
\section{Numerical Examples}
In the following we are going to present numerical results achieved by the presented algorithms\footnote{Code: \href{https://github.com/janhauffen/Block-ALISTA}{https://github.com/janhauffen/Block-ALISTA}}. We will investigate two MMV scenarios. Firstly the measurements $y^l$ are obtained with a random Gaussian matrix and secondly obtained by a redcued-rank random circular convolution. In each scenario we will enforce the Kronecker structure and thus reducing the training cost by training only low dimensional $m\times n$ matrices. Furthermore in the convolutional setting the circular structure will be also enforced, thus reducing the training costs even more. 
To have a fair comparison, all algorithms are initialized with the analytical weight matrix.
\subsection{MMV Setting}
\begin{table}[h]
    \centering
    \begin{tabular}{c|ccc}
        Case & Dimensions & $rank(K)$ & $\mu(K)$ \\ \hline\hline\\
        Gauss & $K\in\R^{32\times 128}$ & 32  & 0.6268\\
        Circ. & $K\in\R^{128\times 128}$ & $32$ & 0.6833
    \end{tabular}
    \caption{Properties of matrix $K$ in both scenarios.}\label{tbl::ProblemInfo}
\end{table}
The training data is sampled from an unknown distribution $\mathcal{X}$ and generated as follows. The signals $x$ are generated for a given number of blocks $n$, given block length $d$ and a possibility if a block $x[i]$ is active or not, i.e., if $\|x[i]\|_2\neq 0$ or $\|x[i]\|_2 = 0$, called $pnz$ (probability of non-zeros). If a block is active, the elements of this blocks are given by a normal Gaussian distribution with variance $\sigma^2 = 1$. The measurements $y$ are obtained by \eqref{underlyingEq}. Where the elements of $\epsilon$ are given from a normal Gaussian distribution with variance $\sigma^2 = pnz \cdot n_x/n_y \cdot 10^{-SNR_{dB} / 10}$.  $SNR_{dB}$ is the signal to noise ratio given in decibel. We consider the following cases. In each case we generate $x$ with $d=15$, $n = 128, m = 32$ and $pnz=10\%$.\\
\noindent\textbf{Gaussian Measurenment Matrix: }
In the Gaussian setting we sample a $m\times n$ matrix $K$ iid from a Gaussian distribution with variance $\sigma^2=1$. We normalize the columns, s.t. $D$ has orthonormal blocks, as assumed in the beginning.  \\
\noindent\textbf{Circular Convolution Matrix: }
We construct the circular matrix as follows. At first we generate a random iid sampled vector $\tilde{a}$ but set a certain amount of elements to zero. We define $k=\Re\left( F^{-1}\tilde{a}\right)$ and thus we can generate a rank deficient Matrix $D=K\otimes I_d$, where $K=circ(k)$. Thus $y^l$ is obtained through a circular convolution with symmetric $\hat{k}=Fk$. It is important for this section to generate a rank deficient matrix to have compressive observations. Otherwise, we would get a trivial problem if $K$, respectively $D$, has full rank. Computing $D=K\otimes I_d$ yields the desired matrix. The properties of these two matrices can be found in Table \ref{tbl::ProblemInfo}.
\subsection{Discussion}
\begin{figure}%
    \centering
    \subfloat[Gaussian Measurement Problem]{{\includegraphics[width=0.5\textwidth,trim={0 6cm 0 6cm},clip]{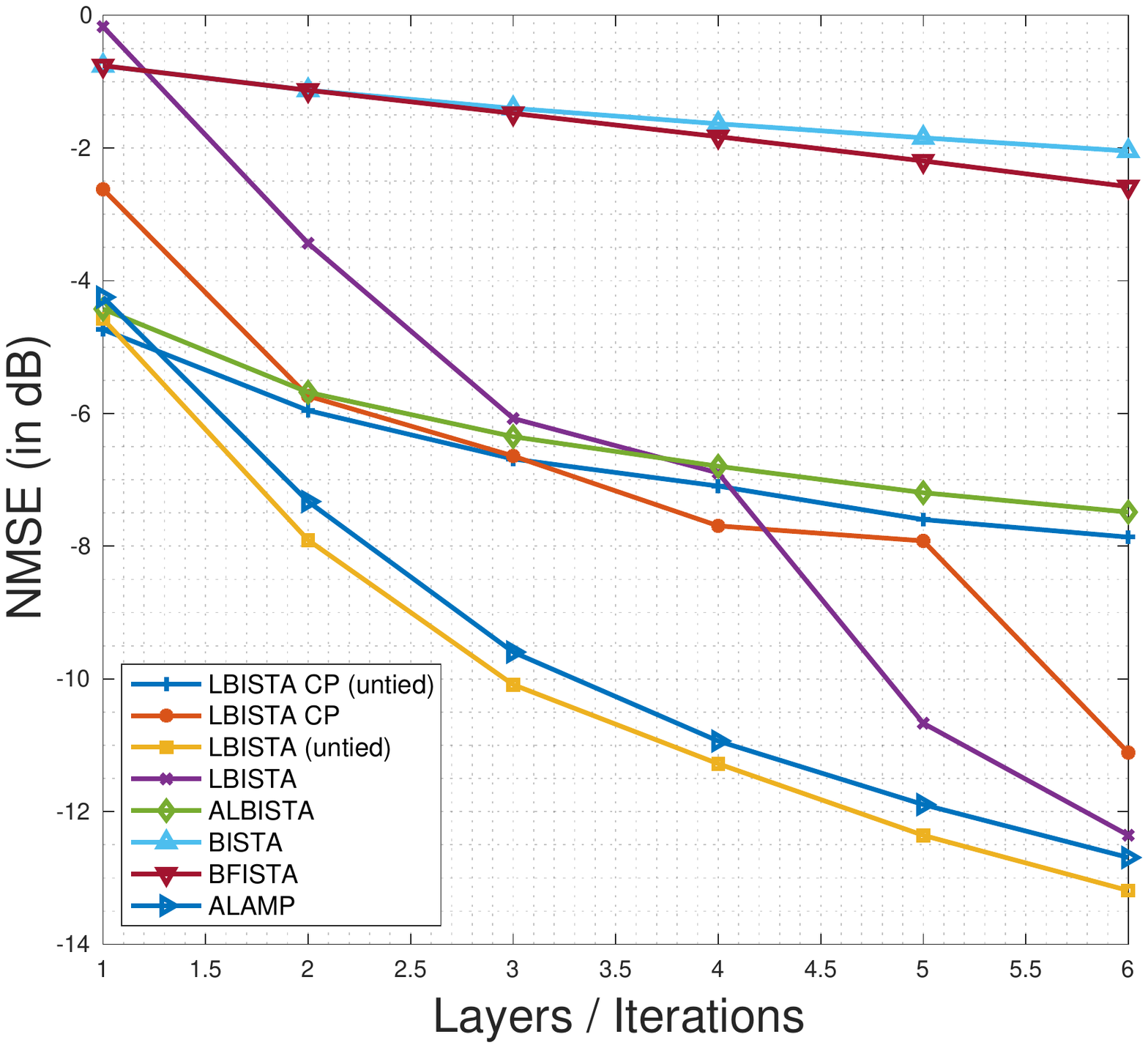} }\label{fig:Gaussian_NMSE_vs_Iter}}%
    \qquad
    \subfloat[Circular Convolution Problem]{{\includegraphics[width=0.5\textwidth,trim={0 6cm 0 6cm},clip]{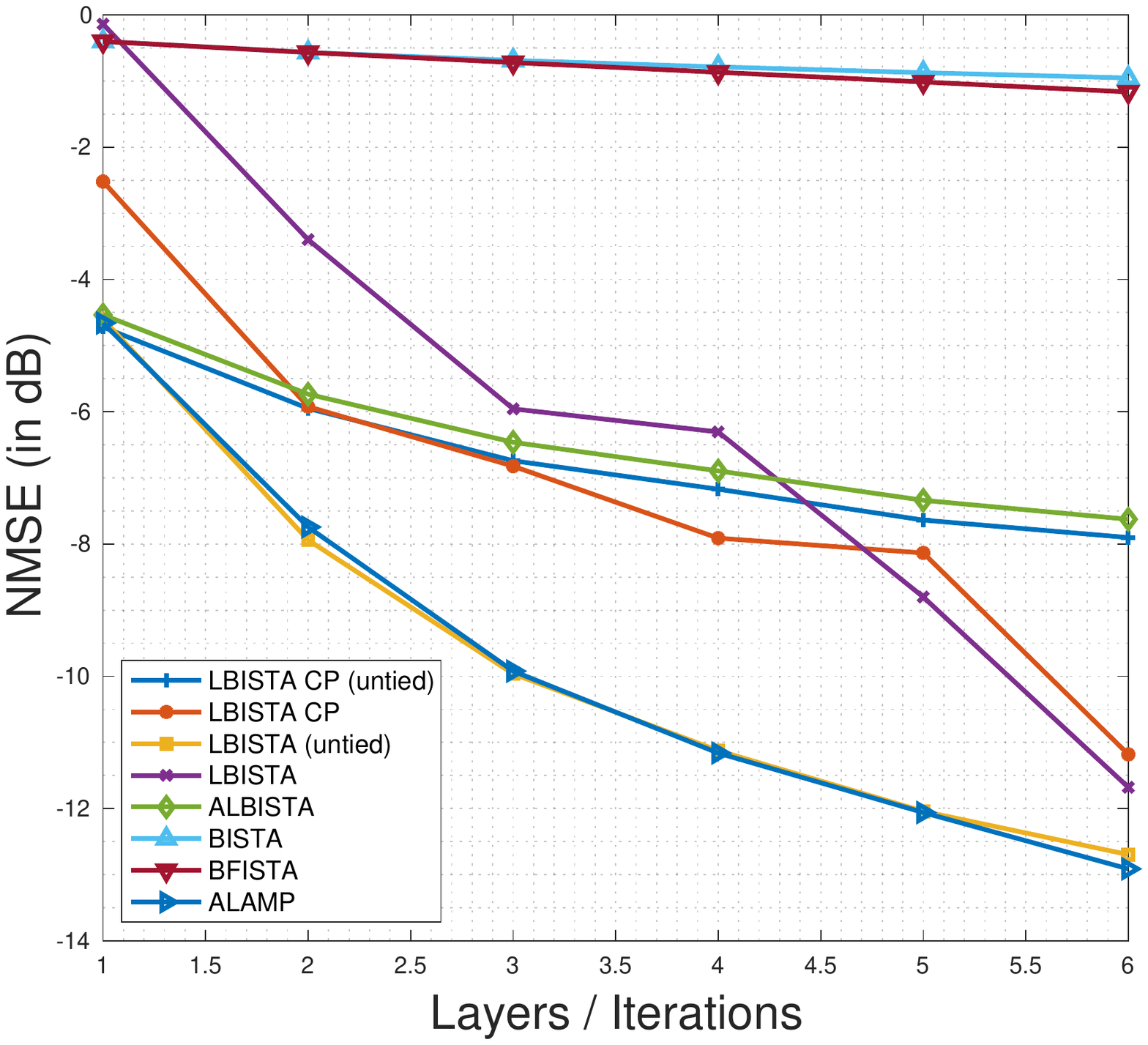}}\label{fig:Conv_NMSE_vs_Iter}}%
    \caption{NMSE in dB over layers / iterations, pnz = 10\%, without noise. BISTA and FastBISTA are evaluated with $\alpha = 1$ and $\gamma = 1 / (1.01\|D\|^2)$.}%
    \label{fig:NMSE_vs_Iter}%
\end{figure}

\begin{figure}%
    \centering
    \subfloat[Parameters and maximal $\ell_{2,1}$-error in the noiseless case.]{{\includegraphics[width=0.5\textwidth,trim={0 6cm 0 6cm},clip]{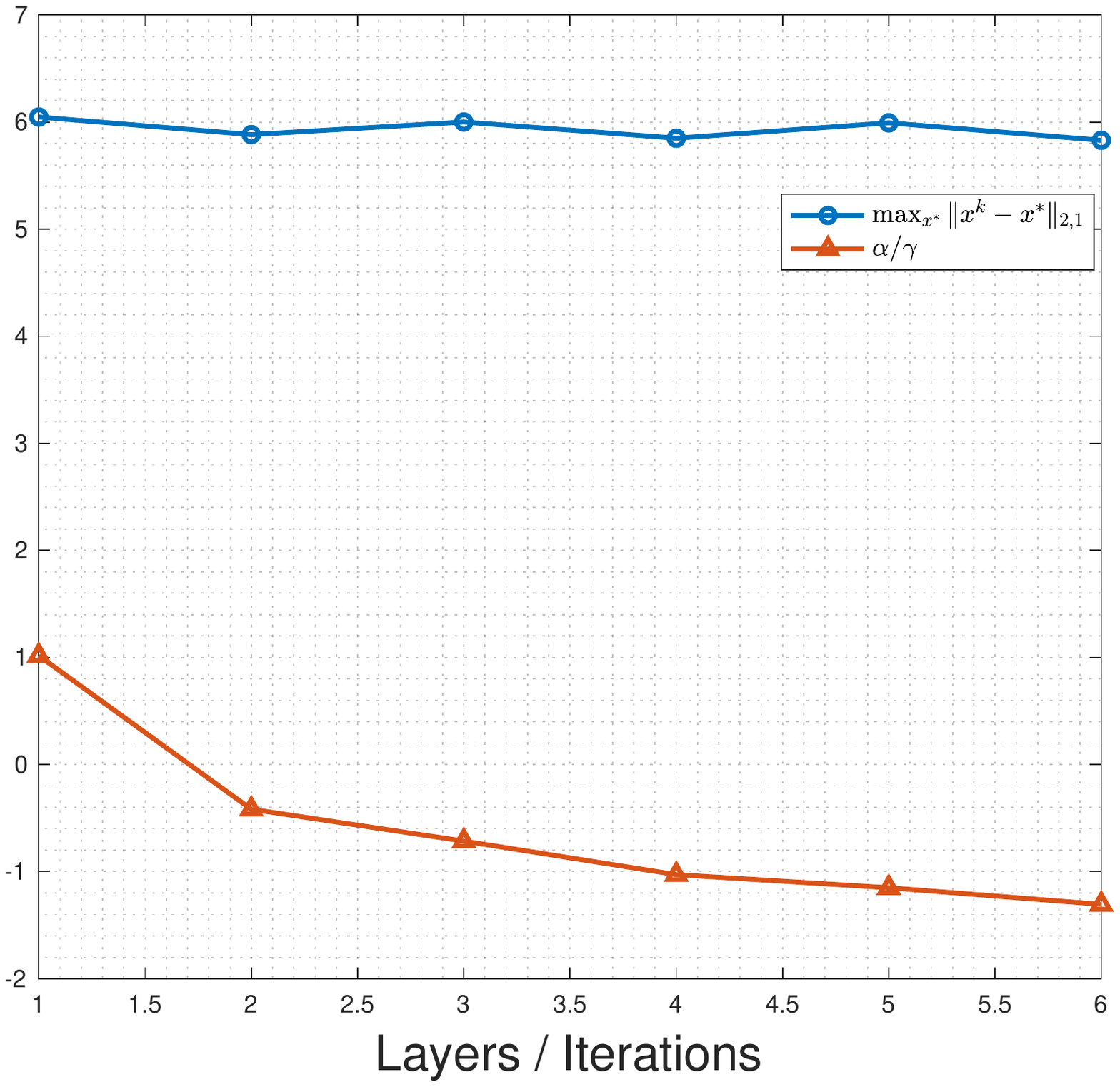} }\label{fig:justification}}%
    \qquad
    \subfloat[Trained $\gamma$ over layers / iterations.]{{\includegraphics[width=0.5\textwidth,trim={0 6cm 0 6cm},clip]{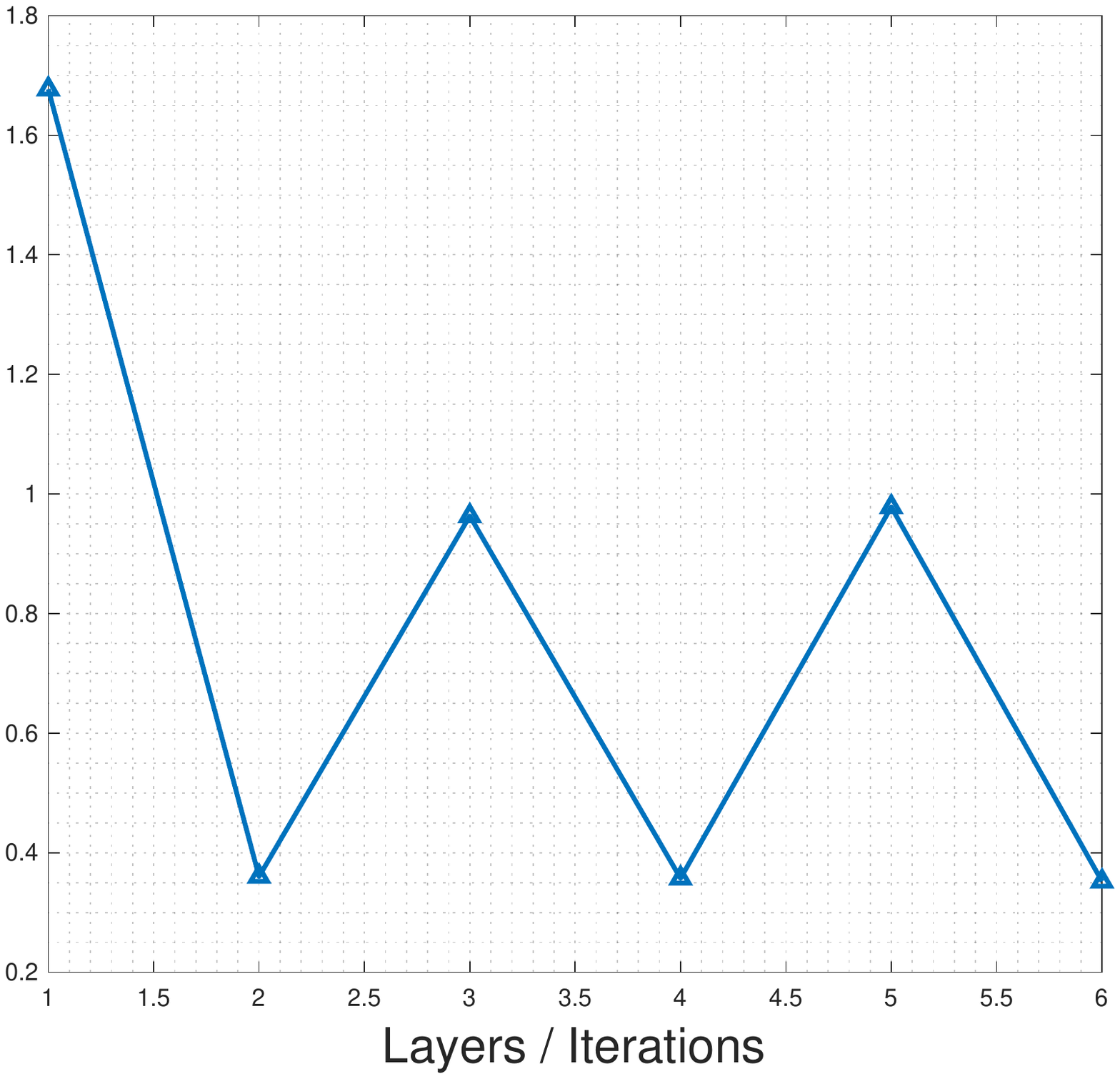}}\label{fig:justification_gam}}%
    \caption{Plots justifying results in Theorem \ref{UpperBound} for the results on the problem with circular convolution matrix in Figure \ref{fig:justification} and with Gaussian measurement matrix ion \ref{fig:justification_gam}. }%
    \label{fig:Thm}%
\end{figure}

\begin{figure}%
    \centering
    \subfloat[Gaussian Measurement Problem]{{\includegraphics[width=0.5\textwidth,trim={0 6cm 0 6cm},clip]{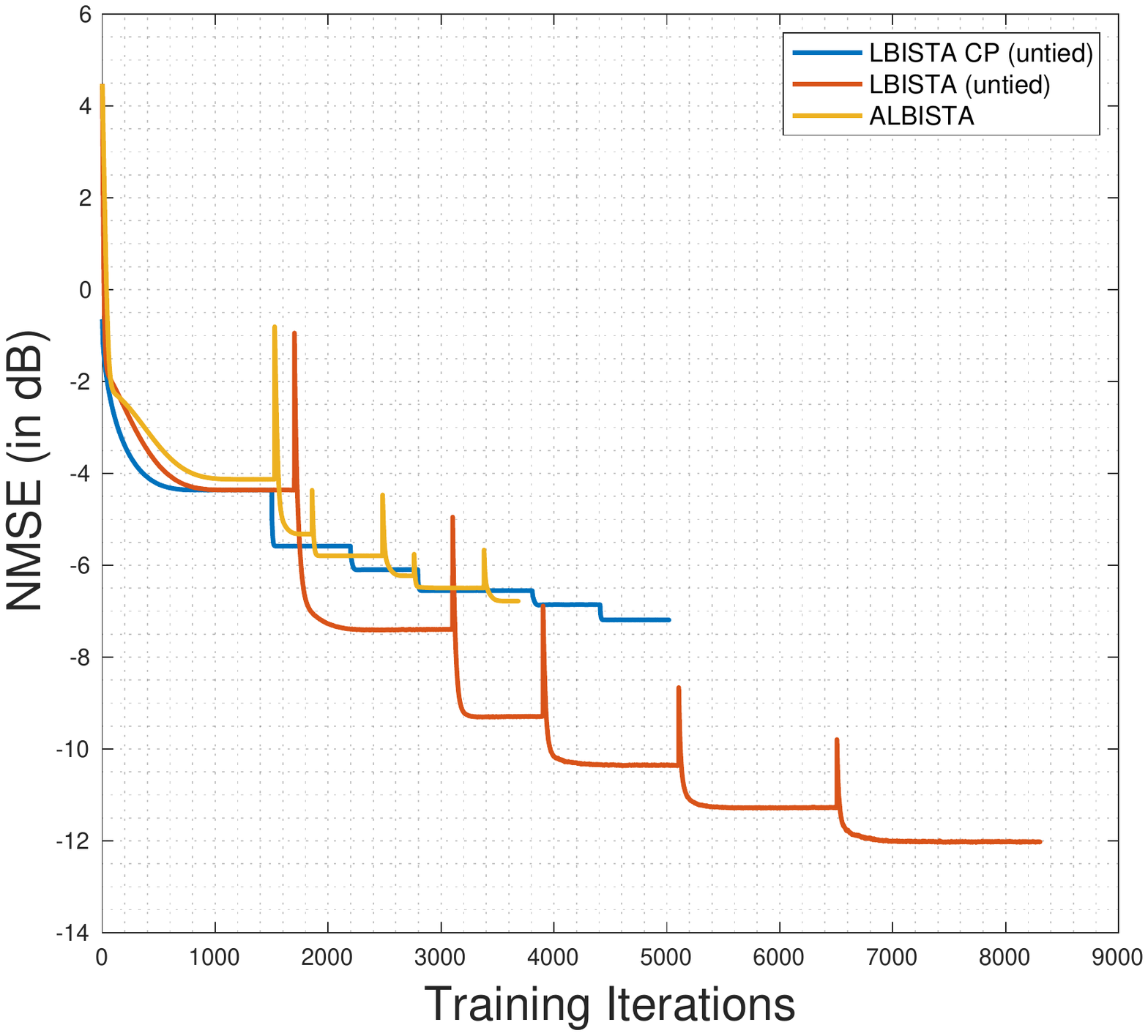} }\label{fig:his_gauss}}%
    \qquad
    \subfloat[Circular Convolution Problem]{{\includegraphics[width=0.5\textwidth,trim={0 6cm 0 6cm},clip]{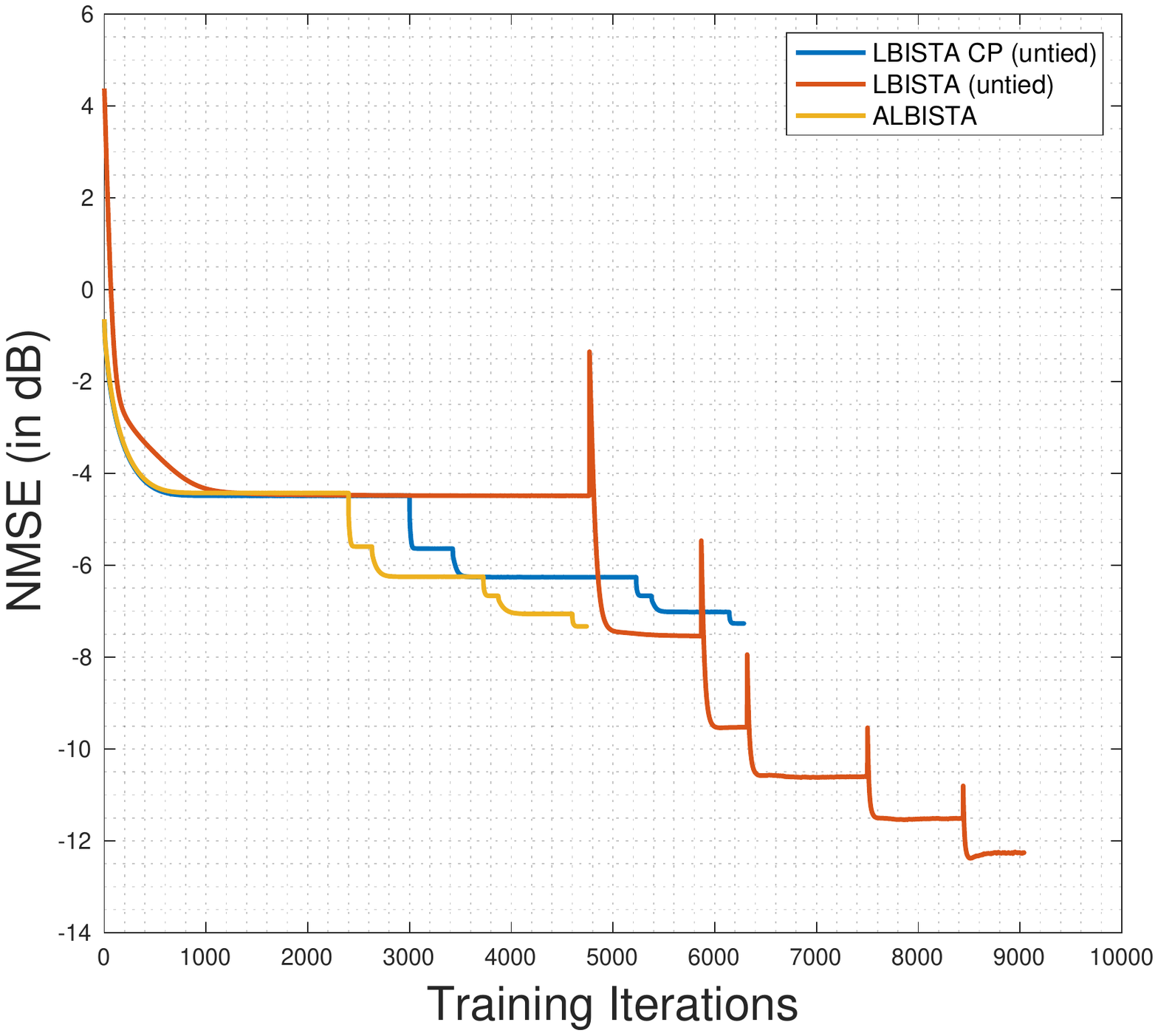}}\label{fig:justification_gam}}%
    \caption{Training history for results shown in Figure \ref{fig:Gaussian_NMSE_vs_Iter} respectively \ref{fig:Conv_NMSE_vs_Iter}. One can see that ALBISTA needs less training iterations than LBISTA CP (untied) or LBISTA (untied).}%
    \label{fig:History}%
\end{figure}

The results of the proposed methods can be found in Figure \ref{fig:Gaussian_NMSE_vs_Iter} for the Gaussian Problem case in Figure \ref{fig:Conv_NMSE_vs_Iter} for the circular case. We also consider the performance of a version of AMP \cite{donoho2009message}. AMP can be viewed as an Bayesian extension of ISTA with an additional Onsager correction term, before applying the thresholding operator, i.e. 
\begin{align*}
    v^{(k)} &  = y - Dx^{(k)} + b^{(k)} v^{(k-1)}\\
    x^{(k+1)} & = \eta_{\alpha }\left(x^{(k)}+\gamma D^Tv^{(k)}\right)
\end{align*}
where $b^{(k)} = \mathbb{E}\left[\eta'(x^{(k)})\right]$. Here we will train only $\gamma, \alpha$, with the same procedure as already discussed and choose $D^T=B^T$. By using $B^T$ instead of $D^T$ we are resembling Orthogonal AMP (OAMP) \cite{ma2017orthogonal}, which follows a similar idea as ALISTA. In \cite{ma2017orthogonal} $B$ is chosen to be de-correlated with respect to $K$, i.e. 
\begin{align*}
   \text{tr}(I_{n_y}-B^TD) = 0 .
\end{align*}
The analytical weight matrix $B$ satisfies also this condition. Moreover in \cite{ma2017orthogonal} the choice of different matrices is also discussed. Thus, untrained ALISTA with correction term can be viewed as a special case of OAMP. Note also, that the structure of Trainable ISTA, proposed in \cite{8695874}, is also based on OAMP and thus there are interrelations between AMP, Unfolding ISTA and Analytical ISTA. We refer to learned AMP with analytical matrix as ALAMP. Differently to \cite{ma2017orthogonal} we use the $\ell_{2,1}$-regularizer, instead of only $\ell_1$-regularization, since we consider the MMV setting. This is also discussed in \cite{kim2011belief, Chen_2018}.
Every proposed algorithm performs better, in terms of \textit{NMSE}, as their untrained original. Interestingly learned AMP, with analytical weight matrix $B$, has a performance almost as good as LBISTA (untied) with only a fraction of trainable parameters. This may come from the fact, that block-soft thresholding is not the correct MMSE estimator for the generated signals $x$ and thus the correction of $b^{(k)}$ yields the better estimation for $x$. As expected we get almost the same performance of ALBISTA and LBISTA CP (untied). In Figure \ref{fig:justification} we show similar plots for the justification of Theorem \ref{UpperBound}, as also seen in \cite{ALISTA}. In particular, Figure \ref{fig:justification} shows that $\frac{\alpha^{(k)}}{\gamma^{(k)}}$ is proportional to the maximal $\ell_{2,1}$-error over all training signals. An interesting behaviour, which carries over from the sparse case, is that the learned $\ell_{2,1}$-regularization parameters approach zero, as $k$ increases.  If $\alpha$ is close to zero, we approach a least-squares problem. This means that after LBISTA found the support of the unknown signal $x^*$ the algorithm consist only of the least squares fitting. Figure \ref{fig:justification_gam} shows that the trained $\gamma^{(k)}$ are bound in an interval. Note that, in contrast to \cite{ALISTA} Theorem \ref{UpperBound} is based on a more general assumption, onto the thresholding parameters. One can take a suitable $\kappa$ and obtains the upper bound for the $\ell_2$-error and thus have convergence on the training set, if the sparsity assumptions are met. 
Figure \ref{fig:History} shows the training loss over the training iterations for the results presented in \ref{fig:NMSE_vs_Iter}. One can see, that ALBISTA needs less training iterations as LBISTA CP (untied) or LBISTA (untied) and thus less training data. The observed jumps occur when moving from one layer to the next, due to the layer-wise training, Algorithm \ref{Training}. A layer is defined to be optimized, if NMSE converged within $1e-5$.
\section{Conclusion}
We proposed ALISTA for the block-sparse and MMV case, important for many real-world applications, and derived corresponding theoretical convergence and recovery results. We relaxed the conditions for the regularization parameter and thus obtained a more precise upper-bound after ALISTA is trained. Nevertheless, this is still dependent on a sharp sparsity assumption on the unknown signals. We investigated and derived a direct solution for the analytical weight matrix in the general block sparse setting as well for one convolutional scenarios. The last section provides numerical results and includes interrelations to AMP.
\clearpage
{\footnotesize
  \bibliographystyle{abbrv}
  \bibliography{bib}

\begin{thebibliography}{10}

\bibitem{ahmadi2020photothermal}
S.~Ahmadi, P.~Burgholzer, G.~Mayr, P.~Jung, G.~Caire, and M.~Ziegler.
\newblock Photothermal super resolution imaging: A comparison of different
  thermographic reconstruction techniques.
\newblock {\em NDT \& E International}, 111:102228, 2020.

\bibitem{bauschke2011convex}
H.~H. Bauschke, P.~L. Combettes, et~al.
\newblock {\em Convex analysis and monotone operator theory in Hilbert spaces},
  volume 408.
\newblock Springer, 2011.

\bibitem{Beck}
A.~Beck.
\newblock { A Fast Iterative Shrinkage-Thresholding Algorithm for Linear
  Inverse Problem}.
\newblock {\em {Society for Industrial and Applied Mathematic}}, 2009.

\bibitem{ben2003generalized}
A.~Ben-Israel and T.~N. Greville.
\newblock {\em Generalized inverses: theory and applications}, volume~15.
\newblock Springer Science \& Business Media, 2003.

\bibitem{bertsekas2014constrained}
D.~P. Bertsekas.
\newblock {\em Constrained optimization and Lagrange multiplier methods}.
\newblock Academic press, 2014.

\bibitem{byrne2008applied}
C.~L. Byrne.
\newblock {\em Applied iterative methods}.
\newblock AK Peters Wellesley, MA, 2008.

\bibitem{candes2006robust}
E.~J. Cand{\`e}s, J.~Romberg, and T.~Tao.
\newblock Robust uncertainty principles: Exact signal reconstruction from
  highly incomplete frequency information.
\newblock {\em IEEE Transactions on information theory}, 52(2):489--509, 2006.

\bibitem{candes2005decoding}
E.~J. Candes and T.~Tao.
\newblock Decoding by linear programming.
\newblock {\em IEEE transactions on information theory}, 51(12):4203--4215,
  2005.

\bibitem{MMVRecovery}
J.~Chen and X.~Huo.
\newblock Theoretical results on sparse representations of multiple-measurement
  vectors.
\newblock {\em IEEE Transactions on Signal Processing}, 54(12):4634--4643,
  2006.

\bibitem{chen2001atomic}
S.~S. Chen, D.~L. Donoho, and M.~A. Saunders.
\newblock Atomic decomposition by basis pursuit.
\newblock {\em SIAM review}, 43(1):129--159, 2001.

\bibitem{ConvUnfoldedISTA}
X.~Chen, J.~Liu, Z.~Wang, and W.~Yin.
\newblock {Theoretical Linear Convergence of Unfolded ISTA and its Practical
  Weights and Thresholds}.
\newblock {\em {conference on Neural Information Processing Systems (NeurIPS
  2018)}}, 2018.

\bibitem{chen2021hyperparameter}
X.~Chen, J.~Liu, Z.~Wang, and W.~Yin.
\newblock Hyperparameter tuning is all you need for lista.
\newblock {\em Advances in Neural Information Processing Systems}, 34, 2021.

\bibitem{Chen_2018}
Z.~Chen, F.~Sohrabi, and W.~Yu.
\newblock Sparse activity detection for massive connectivity.
\newblock {\em {IEEE} Transactions on Signal Processing}, 66(7):1890--1904, apr
  2018.

\bibitem{chitsaz2020acceleration}
K.~Chitsaz, M.~Hajabdollahi, N.~Karimi, S.~Samavi, and S.~Shirani.
\newblock Acceleration of convolutional neural network using fft-based split
  convolutions.
\newblock {\em arXiv preprint arXiv:2003.12621}, 2020.

\bibitem{combettes2011proximal}
P.~L. Combettes and J.-C. Pesquet.
\newblock Proximal splitting methods in signal processing.
\newblock In {\em Fixed-point algorithms for inverse problems in science and
  engineering}, pages 185--212. Springer, 2011.

\bibitem{donoho2006compressed}
D.~L. Donoho.
\newblock Compressed sensing.
\newblock {\em IEEE Transactions on information theory}, 52(4):1289--1306,
  2006.

\bibitem{donoho2005stable}
D.~L. Donoho, M.~Elad, and V.~N. Temlyakov.
\newblock Stable recovery of sparse overcomplete representations in the
  presence of noise.
\newblock {\em IEEE Transactions on information theory}, 52(1):6--18, 2005.

\bibitem{donoho2009message}
D.~L. Donoho, A.~Maleki, and A.~Montanari.
\newblock Message-passing algorithms for compressed sensing.
\newblock {\em Proceedings of the National Academy of Sciences},
  106(45):18914--18919, 2009.

\bibitem{figueiredo2007gradient}
M.~A. Figueiredo, R.~D. Nowak, and S.~J. Wright.
\newblock Gradient projection for sparse reconstruction: Application to
  compressed sensing and other inverse problems.
\newblock {\em IEEE Journal of selected topics in signal processing},
  1(4):586--597, 2007.

\bibitem{fornasier2008iterative}
M.~Fornasier and H.~Rauhut.
\newblock Iterative thresholding algorithms.
\newblock {\em Applied and Computational Harmonic Analysis}, 25(2):187--208,
  2008.

\bibitem{foucart2017mathematical}
S.~Foucart and H.~Rauhut.
\newblock A mathematical introduction to compressive sensing.
\newblock {\em Bull. Am. Math}, 54(2017):151--165, 2017.

\bibitem{fu2021deep}
R.~Fu, V.~Monardo, T.~Huang, and Y.~Liu.
\newblock Deep unfolding network for block-sparse signal recovery.
\newblock In {\em ICASSP 2021-2021 IEEE International Conference on Acoustics,
  Speech and Signal Processing (ICASSP)}, pages 2880--2884. IEEE, 2021.

\bibitem{gorodnitsky1995neuromagnetic}
I.~F. Gorodnitsky, J.~S. George, and B.~D. Rao.
\newblock Neuromagnetic source imaging with focuss: a recursive weighted
  minimum norm algorithm.
\newblock {\em Electroencephalography and clinical Neurophysiology},
  95(4):231--251, 1995.

\bibitem{gregor2010learning}
K.~Gregor and Y.~LeCun.
\newblock Learning fast approximations of sparse coding.
\newblock In {\em Proceedings of the 27th international conference on
  international conference on machine learning}, pages 399--406, 2010.

\bibitem{hung1975moore}
C.-h. Hung and T.~L. Markham.
\newblock The moore-penrose inverse of a partitioned matrix m=(adbc).
\newblock {\em Linear Algebra and its Applications}, 11(1):73--86, 1975.

\bibitem{8695874}
D.~Ito, S.~Takabe, and T.~Wadayama.
\newblock Trainable ista for sparse signal recovery.
\newblock {\em IEEE Transactions on Signal Processing}, 67(12):3113--3125,
  2019.

\bibitem{ALISTA}
Z.~W. Jialin~Liu, Xiaohan~Chen and W.~Yin.
\newblock {ALISTA: ANALYTIC WEIGHTS ARE AS GOOD AS LEARNED WEIGHTS IN LISTA}.
\newblock {\em {conference paper at ICLR}}, 2019.

\bibitem{kim2020element}
D.~Kim and D.~Park.
\newblock Element-wise adaptive thresholds for learned iterative shrinkage
  thresholding algorithms.
\newblock {\em IEEE Access}, 8:45874--45886, 2020.

\bibitem{kim2011belief}
J.~Kim, W.~Chang, B.~Jung, D.~Baron, and J.~C. Ye.
\newblock Belief propagation for joint sparse recovery, 2011.

\bibitem{kingma2017adam}
D.~P. Kingma and J.~Ba.
\newblock Adam: A method for stochastic optimization, 2015.

\bibitem{kutyniok2014compressed}
G.~Kutyniok.
\newblock Compressed sensing.
\newblock {\em Mitteilungen der Deutschen Mathematiker-Vereinigung}, 1:24--29,
  2014.

\bibitem{ma2017orthogonal}
J.~Ma and L.~Ping.
\newblock Orthogonal amp.
\newblock {\em IEEE Access}, 5:2020--2033, 2017.

\bibitem{musa2021plug}
O.~Musa, P.~Jung, and G.~Caire.
\newblock Plug-and-play learned gaussian-mixture approximate message passing.
\newblock In {\em ICASSP 2021-2021 IEEE International Conference on Acoustics,
  Speech and Signal Processing (ICASSP)}, pages 4855--4859. IEEE, 2021.

\bibitem{pratt2017fcnn}
H.~Pratt, B.~Williams, F.~Coenen, and Y.~Zheng.
\newblock Fcnn: Fourier convolutional neural networks.
\newblock In {\em Joint European Conference on Machine Learning and Knowledge
  Discovery in Databases}, pages 786--798. Springer, 2017.

\bibitem{rudelson2005geometric}
M.~Rudelson and R.~Vershynin.
\newblock Geometric approach to error-correcting codes and reconstruction of
  signals.
\newblock {\em International mathematics research notices},
  2005(64):4019--4041, 2005.

\bibitem{schacke2004kronecker}
K.~Schacke.
\newblock On the kronecker product.
\newblock {\em Master's thesis, University of Waterloo}, 2004.

\bibitem{shannon1949communication}
C.~E. Shannon.
\newblock Communication in the presence of noise.
\newblock {\em Proceedings of the IRE}, 37(1):10--21, 1949.

\bibitem{Eldar}
H.~B. Yonina C.~Eldar.
\newblock { BLOCK-SPARSITY: COHERENCE AND EFFICIENT RECOVERY}.
\newblock {\em {arXiv:0812.0329}}, 2008.

\bibitem{ziniel2012efficient}
J.~Ziniel and P.~Schniter.
\newblock Efficient high-dimensional inference in the multiple measurement
  vector problem.
\newblock {\em IEEE Transactions on Signal Processing}, 61(2):340--354, 2012.

\end{thebibliography}
}
\onecolumn
\begin{appendices}
\section{Proof of Theorem \ref{UpperBound}}
The following proof combines proofs from \cite{ALISTA} and \cite{ConvUnfoldedISTA} for the block-sparse setting with additional noise.
\begin{proof}
We proof (\ref{support2}) by induction: With $k=0$ this statement is satisfied, because $x^{(0)}=0$. \\
We fix $k$ and assume (\ref{support2}) to be true.
\begin{align*}
x^{(k+1)}[i]&=\eta_{\alpha^{(k)}} \left( x^{(k)}[i]-\gamma^{(k)}B[i]^T(Dx^{(k)}-y)\right) \\
&=\eta_{\alpha^{(k)}} \left( x^{(k)}[i]-\gamma^{(k)}\left(B[i]^TD(x^{(k)}-x^*)-B[i]^T\epsilon\right)\right) \\
&=\eta_{\alpha^{(k)}} \left( -\gamma^{(k)}\left(\sum_{j\in\mathbb{S}}B[i]^TD[j](x^{(k)}[j]-x^*[j])-B[i]^T\epsilon\right)\right) \forall i\not\in\mathbb{S}
\end{align*}
From \eqref{eq:thm1:alphak} in the theorem, we have for all $i\not\in\mathbb{S}$ 
\begin{align*}
\alpha^{(k)}& \geq d\gamma^{(k)}\tilde{\mu}_b \|x^{(k)}-x^*\|_{2,1}+C\sigma\\
& \geq d\gamma^{(k)}\tilde{\mu}_b \|x^{(k)}-x^*\|_{2,1}+C\|\epsilon\|_2\\
& = \gamma^{(k)}\sum_{j=1}^M d \tilde{\mu}_b\|x^{(k)}[j]-x^*[j]\|_{2}+C\|\epsilon\|_2 \\
& =\gamma^{(k)} \sum_{j\in \text{supp}_b(x^{(k)})} d \tilde{\mu}_b\|x^k[j]-x^*[j]\|_{2} +C\|\epsilon\|_2\\
& =\gamma^{(k)} \sum_{j\in \mathbb{S}}\tilde{\mu}_b\|x^{(k)}[j]-x^*[j]\|_{2} +C\|\epsilon\|_2\\
& \geq  \gamma^{(k)}\sum_{j\in \mathbb{S}}\left(\|B[i]^TD[j]\|_2\|x^{(k)}[j]-x^*[j]\|_{2} + \|B[i]\|_2 \|\epsilon\|_2\right).
\end{align*}
From this we get the following inequality
\begin{align*}
\| -\gamma^{(k)}\left(\sum_{j\in\mathbb{S}}B[i]^TD[j](x^{(k)}[j]-x^*[j])+B[i]^T\epsilon\right) \|_2 \\ \leq \gamma^{(k)} \left(\sum_{j\in\mathbb{S}}\|B[i]^TD[j](x^{(k)}[j]-x^*[j])\|_2+\|B[i]^T\epsilon\|_2\right) \\
 \leq \gamma^{(k)} \left(\sum_{j\in\mathbb{S}}\|B[i]^TD[j]\|_2\|(x^{(k)}[j]-x^*[j])\|_2+\|B[i]^T\epsilon\|_2\right) \\
 \leq \gamma^{(k)} \left(\sum_{j\in\mathbb{S}}\|B[i]^TD[j]\|_2\|(x^{(k)}[j]-x^*[j])\|_2+\|B[i]\|_2 \| \epsilon\|_2\right)\leq\alpha^{(k)}
\end{align*}
meaning that we get for all $i\not\in\mathbb{S}$: 
\begin{align*}
\frac{\alpha^{(k)}}{\| -\gamma^{(k)}\sum_{j\in\mathbb{S}}B[i]^TD[j](x^{(k)}[j]-x^*[j])+B[i]^T\epsilon\|_2} &\geq 1.
\end{align*}
implying for the thresholding operator $\eta_{\alpha^{(k)}}$ that $x^{(k+1)}[i]=0$, $\forall i \notin \mathbb{S}$, and with this (\ref{support2}) is proven by induction.

We now show (\ref{upperbound2}). Take $(x^*, \epsilon)\in\mathcal{X}_b(M,s, \sigma)$. For all $i\in\mathbb{S}$ we get
\begin{align*}
x^{(k+1)}[i]&=\eta _{\alpha^{(k)}}\left(x^{(k)}[i]-\gamma^{(k)}B[i]^TD[(j\in\mathbb{S})]\left(x^{(k)}[(j\in\mathbb{S})]-x^*[(j\in\mathbb{S})]\right) +\gamma^{(k)}B[i]^T\epsilon \right)\\
& \in x^{(k)}[i]-\gamma^{(k)}B[i]^TD[(j\in\mathbb{S})]\left(x^{(k)}[(j\in\mathbb{S})]-x^*[(j\in\mathbb{S})]\right)+\gamma^{(k)}B[i]^T\epsilon-\alpha^{(k)} \partial \ell_2(x^{(k+1)}[i])
\end{align*}
where $\partial\ell_2 $ is the sub-gradient of $\|x\|_2$, with $x\in\R^{n_x}$:
\begin{align*}
\partial \ell_2(x)=
\left\{ \begin{tabular}{lc}
$B_1^2(0)$ &, $x=0$ \\
$\left\lbrace \left( \frac{|x_i|sign(x_i)}{\|x\|_2}  \right)_{i=1}^{n_x}\right\rbrace $&,  $x\neq$ 0
\end{tabular}  \right .
\end{align*}
where $B_1^2(0)$ is the ball with radius 1 in the $\ell_2$-norm around $0$. With the choice of $B\in\mathcal{W}_b(D)$ we know $B[i]^TD[i]=I_d$ for all $i$.
\begin{align*}
\Rightarrow & x^{(k)}[i]-\gamma^{(k)}B^T[i]D[(j\in\mathbb{S})]\left(x^{(k)}[(j\in\mathbb{S})]-x^*[(j\in\mathbb{S})]\right) \\
&= x^{(k)}[i]-\gamma^{(k)}\sum_{j\in\mathbb{S}, j\neq i}B[i]^TD[j]\left(x^{(k)}[j]-x^*[j]\right) - \gamma^{(k)}\left(x^{(k)}[i]-x^*[i] \right)\\
& = x^*[i]-\gamma^{(k)}\sum_{j\in\mathbb{S}, j\neq i}B[i]^TD[j]\left(x^{(k)}[j]-x^*[j]\right)+\left(1 - \gamma^{(k)}\right)\left(x^{(k)}[i]-x^*[i] \right).
\end{align*}
Then it holds for all $i \in\mathbb{S}$
\begin{align*}
x^{(k+1)}[i]-x^*[i] \in &-\gamma^{(k)}\sum_{j\in\mathbb{S}, j\neq i}B^T[i]D[j]\left(x^{(k)}[j]-x^*[j]\right)+\left(1 - \gamma^{(k)}\right)\left(x^{(k)}[i]-x^*[i] \right)\\&+\gamma^{(k)}B[i]^T\epsilon-\alpha^{(k)} \partial \ell_2(x^{(k+1)}[i])
\end{align*}
We know that for all $x\in\partial \ell_2(x)$ it holds that $\|x\|_2\leq 1$.
\begin{align*}
\Rightarrow &\|x^{(k+1)}[i]-x^*[i]\|_2 \\& \leq \gamma^{(k)}\sum_{j\in\mathbb{S}, j\neq i}\|B^T[i]D[j]\left(x^{(k)}[j]-x^*[j]\right)\|_2+|1 - \gamma^{(k)}|\|x^{(k)}[i]-x^*[i] \|_2+\|\gamma^{(k)}B[i]^T\epsilon\|_2+\alpha^{(k)} \\
& \leq \gamma^{(k)}\sum_{j\in\mathbb{S}, j\neq i}\|B^T[i]D[j]\|_2\|x^{(k)}[j]-x^*[j]\|_2+|1 - \gamma^{(k)}|\|x^{(k)}[i]-x^*[i] \|_2+C\sigma+\alpha^{(k)} \\
& \leq \gamma^{(k)}\sum_{j\in\mathbb{S}, j\neq i}d\tilde{\mu}_b\|x^{(k)}[j]-x^*[j]\|_2+|1 - \gamma^{(k)}|\|x^{(k)}[i]-x^*[i] \|_2+C\sigma+\alpha^{(k)}\\
&  = d\tilde{\mu}_b\gamma^{(k)}\sum_{j\in\mathbb{S}, j\neq i}\|x^{(k)}[j]-x^*[j]\|_2+|1 - \gamma^{(k)}|\|x^{(k)}[i]-x^*[i] \|_2+C\sigma+\alpha^{(k)}
\end{align*}
By equation \ref{support2} we know $\|x^{(k)} - x^* \|_{2,1} = \|x^{(k)}[j\in\mathbb{S}] - x^*[j\in\mathbb{S}] \|_{2,1}$ from this we get: 
\begin{align*}
\|x^{(k+1)} - x^*  \|_{2,1} & = \sum_{i\in\mathbb{S}} \| x^{(k)}[i] - x^*[i]\|_2 \\
& \leq \sum_{i\in\mathbb{S}} \left(d\tilde{\mu}_b\gamma^{(k)}\sum_{j\in\mathbb{S}, j\neq i}\|x^{(k)}[j]-x^*[j]\|_2+|1 - \gamma^{(k)}|\|x^{(k)}[i]-x^*[i] \|_2+C\sigma+\alpha^{(k)} \right) \\
& = d\tilde{\mu}_b\gamma^{(k)}\sum_{i\in\mathbb{S}}\sum_{j\in\mathbb{S}, j\neq i}\|x^{(k)}[j]-x^*[j]\|_2+|1 - \gamma^{(k)}|\sum_{i\in\mathbb{S}}\|x^{(k)}[i]-x^*[i] \|_2+C\sigma+\alpha^{(k)} \\
& = d\tilde{\mu}_b\gamma^{(k)}\left(|\mathbb{S}|-1\right)\sum_{i\in\mathbb{S}}\|x^{(k)}[i]-x^*[i]\|_2+|1 - \gamma^{(k)}|\|x^{(k)}-x^* \|_{2,1}+|\mathbb{S}|(\alpha^{(k)}+C\sigma)\\
& = d\tilde{\mu}_b\gamma^{(k)}\left(|\mathbb{S}|-1\right)\|x^{(k)}[i]-x^*[i]\|_{2,1}+|1 - \gamma^{(k)}|\|x^{(k)}-x^* \|_{2,1}+|\mathbb{S}|(\alpha^{(k)}+C\sigma) \\ 
& = \left(d\tilde{\mu}_b\gamma^{(k)}\left(|\mathbb{S}|-1\right)+|1 - \gamma^{(k)}|\right)\|x^{(k)}-x^* \|_{2,1}+|\mathbb{S}|(\alpha^{(k)}+C\sigma)
\end{align*}
We take the supremum over $(x^*,\epsilon)\in\mathcal{X}_b(M,s,\sigma)$ of the previous inequality and get
\begin{align*}
\sup_{(x^*,\epsilon)}& \left\lbrace \|x^{(k+1)}-x^* \|_{2,1} \right\rbrace \leq \left(d\tilde{\mu}_b\gamma^{(k)}\left(s-1\right)+|1 - \gamma^{(k)}|\right)\sup_{(x^*,\epsilon)}\left\lbrace\|x^{(k)}-x^* \|_{2,1}\right\rbrace+s(\alpha^{(k)}+C\sigma) \\
& \leq \left(d\tilde{\mu}_b\gamma^{(k)}\left(s-1\right)+|1 - \gamma^{(k)}|\right)\sup_{(x^*,\epsilon)}\left\lbrace\|x^{(k)}-x^* \|_{2,1}\right\rbrace+s(\kappa d\gamma^{(k)}\tilde{\mu}_b\sup_{(x^*,\epsilon)} \lbrace \|x^{(k)}-x^*\|_{2,1} \rbrace +C\sigma) \\
& =  \left(d\tilde{\mu}_b\gamma^{(k)}\left(s-1\right)+|1 - \gamma^{(k)}|+\kappa d\gamma^{(k)}\tilde{\mu}_bs\right)\sup_{(x^*,\epsilon)}\left\lbrace\|x^{(k)}-x^* \|_{2,1}\right\rbrace+sC\sigma \\
& =  \left(d\gamma^{(k)}\left((\kappa+1)\tilde{\mu}_bs-\tilde{\mu}_b\right)+|1 - \gamma^{(k)}|\right)\sup_{(x^*,\epsilon)}\left\lbrace\|x^{(k)}-x^* \|_{2,1}\right\rbrace +sC\sigma
\end{align*}
Let $a(k) = d\gamma^{(k)}\left((\kappa+1)\tilde{\mu}_bs-\tilde{\mu}_b\right)+|1 - \gamma^{(k)}|$. We get
\begin{align*}
\sup_{(x^*,\epsilon)} \left\lbrace \|x^{(k+1)}-x^* \|_{2,1} \right\rbrace & \leq 
a(k)\sup_{(x^*,\epsilon)} \left\lbrace \|x^{(k)}-x^* \|_{2,1} \right\rbrace + sC\sigma \\
& \leq a(k)\left(a(k-1)\sup_{(x^*,\epsilon)} \left\lbrace \|x^{(k-1)}-x^* \|_{2,1} \right\rbrace + sC\sigma  \right) + sC\sigma \\
& = a(k)a(k-1)\sup_{(x^*,\epsilon)} \left\lbrace \|x^{(k-1)}-x^* \|_{2,1} \right\rbrace + s\left(a(k) C\sigma + C\sigma \right)\\
& \leq a(k)a(k-1)\left(a(k-2)\left(\sup_{(x^*,\epsilon)} \left\lbrace \|x^{(k-2)}-x^* \|_{2,1} \right\rbrace + C\sigma\right)  \right) + s\left(a(k) C\sigma + C\sigma \right)\\
& = a(k)a(k-1)a(k-2)\sup_{(x^*,\epsilon)} \left\lbrace \|x^{(k-2)}-x^* \|_{2,1} \right\rbrace + s\left(a(k)a(k-1)C\sigma + a(k) C\sigma + C\sigma \right) \\
& \vdots \\
& \leq \prod\limits_{\tau = 0}^{k}a(\tau)\sup_{(x^*,\epsilon)}\left\lbrace \|x^{(0)}-x^* \|_{2,1} \right\rbrace + sC\sigma\left(1+\sum_{\tau = 0}^{k}\prod\limits_{s = \tau}^{k}a(s) \right)
\end{align*}
Set $\tilde{a}(k)=-\log ( a(k)) $ then the previous inequality can be reformulated into the following
\begin{align*}
\sup_{(x^*,\epsilon)}\left\lbrace \|x^{(k+1)}-x^* \|_{2,1} \right\rbrace & \leq \exp\left(-\sum\limits_{\tau = 0}^{k}\tilde{a}(\tau)\right)\sup_{(x^*,\epsilon)}\left\lbrace \|x^{(0)}-x^* \|_{2,1} \right\rbrace + C\sigma\left(1+\sum_{\tau = 0}^{k} \exp\left(-\sum\limits_{s = \tau}^{k-\tau}\tilde{a}(s)\right) \right) \\
 & \leq \exp\left(-\sum\limits_{\tau = 0}^{k}\tilde{a}(\tau)\right)sM+ C\sigma\left(1+\sum_{\tau = 0}^{k} \exp\left(-\sum\limits_{s = \tau}^{k-\tau}\tilde{a}(s)\right) \right).
\end{align*}
And with $\| \cdot \|_{2} \leq \|\cdot \|_{2,1}   $ we get the final inequality.\\
From $s<(d+1/\tilde{\mu}_b)/2d$ we get $2ds\tilde{\mu}_b-d\tilde{\mu}_b<1$ and with this we get for $0<\gamma^{(\tau)}\leq 1$ that $\tilde{\alpha}(\tau)>0$. In the other case $1<\gamma^{(\tau)}<\frac{2}{d(2\tilde{\mu}_bs-\tilde{\mu}_b)+1}$ we have
\begin{align*}
d\gamma^{(\tau)}\left(2\tilde{\mu}_Bs-\tilde{\mu}_B\right)+|1 - \gamma^{(k)}|= \gamma^{(\tau)}\left(d\left(2\tilde{\mu}_Bs-\tilde{\mu}_B\right)+1\right)-1<1
\end{align*}
And therefore we have $\tilde{\alpha}(\tau)>0$.
\end{proof}


\section{Proof of Theorem \ref{Thm2:LowerBound}}
The proof of Theorem \ref{Thm2:LowerBound} follows the lines of  \cite{ALISTA} with some minor modifications to adapt to our block-sparse case: 
\begin{itemize}
    \item change $\ell_1$-norm to $\ell_{2,1}$-norm,
    \item change the support of a sparse vector to the block-support used in this work,
    \item change the set of $n$-dimensional sign numbers to the set $\mathcal{S}(n) = \left\lbrace(s_1, s_2, \dots,s_n) \, |\, s_i\in \mathcal{S}_d\cup \lbrace 0 \rbrace\, \forall\, i=1,\dots,n \right\rbrace$ where $S_d$ is the sphere in $\mathbb{R}^d$ with radius $1$.
\end{itemize}
For the sake of completeness and better readability, we summarize the main steps below.

\begin{proof}
By the law of total probability we may write 
\begin{align}\label{eq:total-prob}
P(\eqref{LowerBoundWithoutNoise}\, \text{holds}) &= \sum_{\mathbb{S}, 2\leq |\mathbb{S}|\leq s}P( \eqref{LowerBoundWithoutNoise}\,\text{holds} \, \vert \, \text{supp}_b(x^*) = \mathbb{S})P(\text{supp}_b(x^*) = \mathbb{S}) \, .
\end{align}
Hence, we seek a bound for the conditional probabilities 
\begin{align}
    P( \eqref{LowerBoundWithoutNoise}\,\text{ does not hold} \, \vert \, \text{supp}_b(x^*) = \mathbb{S}) \, .  \label{eq::goal}
\end{align}
To this end, we divide the proof into 2 steps: \\
\textbf{Step 1: } 
Fix $k$ and assume that  \eqref{LowerBoundWithoutNoise} does not hold. To examine which conditions  $x^*$ should satisfy in this case we introduce the sets 
\begin{align*}
\mathcal{X}^{(k)}(\epsilon) = \left\lbrace x^* \, | \,  \|x^{(k)}-x^*\|_{2,1} < \epsilon \|x^*\|_2\exp(-ck)\right\rbrace
\end{align*}
and 
\begin{align*}
\Tilde{\mathcal{X}}^{(k)} (\epsilon)= \left\lbrace x^* \, |\,  \|x^*[i]\|_2\leq\epsilon \|x^*\|_2\exp(-ck) \text{ for some } i\in\mathbb{S}\right\rbrace \, .
\end{align*}
Following  \cite{ALISTA} it can be shown that 
\begin{align}
\mathcal{X}^{(k)}(\epsilon) \subset \Tilde{\mathcal{X}}^{(k)} (\epsilon) \cup \Tilde{\mathcal{X}}^{(k-1)} (\epsilon) \cup\dots\cup\Tilde{\mathcal{X}}^{(1)} (\epsilon) \cup \mathcal{X}^{(0)}(\epsilon)   \label{eq::sets}
\end{align}
with 
\begin{align*}
\mathcal{X}^{(k-j)}(\epsilon) = \left\lbrace x^* \, | \,  \|(x^{(k-j)}[i\in\mathbb{S}]-x^*[i\in\mathbb{S}])- \tilde{x}^{(k-j)}\|_{2,1} \leq  \epsilon\|x^*\|_2 \frac{\bar{\sigma}_{min}^{k-j}}{3^{sk}} \right\rbrace
\end{align*}
and 
\begin{align*}
\Tilde{\mathcal{X}}^{(k-j)} (\epsilon)= \left\lbrace x^* \, |\,  \|x^*[i]+\tilde{x}^{(k-j)}\|_2<\epsilon \|x^*\|_2\frac{\bar{\sigma}_{min}^{k-j}}{3^{sk}} \text{ for some } i\in\mathbb{S}\right\rbrace \, ,
\end{align*}
with bias given as 
\[ \tilde{x}^{(k-j)}=\sum_{t=1}^j\alpha^{(k-j+t-1)}\left(I-\left(B^{(k-j+t-1)}[i\in\mathbb{S}]\right)^TD[i\in\mathbb{S}]\right)^{-1}\frac{x^{(k-j+t)}[i\in\mathbb{S}]}{\|x^{(k-j+t)}[i\in\mathbb{S}]\|_2} \, ,\]  
for $j=0,\dots, k$.
\\
\textbf{Step 2:} We now estimate the probabilities for the sets in \eqref{eq::sets} and show \eqref{eq::goal} by utilizing 
\begin{align*}
\frac{x^{(k)}[i\in\mathbb{S}]}{\|x^{(k)}[i\in\mathbb{S}]\|_2}\in \mathcal{S}_d\cup \lbrace 0 \rbrace \, ,
\end{align*}
where $\mathcal{S}_d$ is the sphere in $\R^d$ with radius 1. We then obtain 
\begin{align*}
P\left(x^*\in\Tilde{\mathcal{X}}^{(k-j)}\, |\, \text{supp}_b(x^*)=\mathbb{S}\right)
&\leq \epsilon |\mathbb{S}|^{3/2}\frac{\bar{\sigma}_{min}^{k-j}}{3^{(j-k)|\mathbb{S}|}}
\end{align*}
and 
\begin{align*}
P&\left(x^*\in\mathcal{X}^{(0)}(\epsilon)\, |\, \text{supp}_b(x^*)=\mathbb{S}\right) \leq \epsilon^{|\mathbb{S}|}\, . 
\end{align*}
Collecting these results, we get
\begin{align*}
P&\left(x^*\in\mathcal{X}^{(k)}(\epsilon)\, |\, \text{supp}_b(x^*)=\mathbb{S}\right)\\
\leq &\sum_{j=0}^{k-1} \epsilon |\mathbb{S}|^{3/2}\frac{\bar{\sigma}_{min}^{k-j}}{3^{(j-k)|\mathbb{S}|}}+\epsilon^{|\mathbb{S}|}\\
= & \epsilon |\mathbb{S}|^{3/2}\frac{\bar{\sigma}_{min}3^{-|\mathbb{S}|}}{1-\bar{\sigma}_{min}3^{-|\mathbb{S}|}}\left(1-(\bar{\sigma}_{min}3^{-|\mathbb{S}|})^k\right)+\epsilon^{|\mathbb{S}|}\\
\leq & \epsilon |\mathbb{S}|^{3/2}+\epsilon^{|\mathbb{S}|}\,.
\end{align*}
The final result then follows from \eqref{eq:total-prob}.
\end{proof}

\section{Proof of Theorem \ref{thm::AnalyticalMatrix-Up}}\label{proof:upperbound}
\begin{proof}
First, we notice
\begin{align*}
\|D^TB \|_F^2 = \sum_{i=1}^n\|D^TB[i] \|_F^2
\end{align*}
therefore we can reduce \eqref{reducedmutblockcoh}:
    \begin{align}\label{eq::reducedreducedmutblock}
        \forall i:\,
        \min_{B[i]} \|D^TB[i]\|_F^2\\
        s.t. \, D[i]^TB[i]=I_d \, .
    \end{align}
This can be solved with the method of Lagrange multipliers, see for example \cite{bertsekas2014constrained}. The Lagrangian function is defined as 
\begin{align*}
L(x,\lambda) = f(x) + \langle \lambda, h(x)\rangle
\end{align*}
where $f$ is the objective function and $h$ are the equality constraints, i.e. minimize $f(x)$ such that $h(x)=0$ and $\langle\cdot, \cdot \rangle$ an appropriate inner product. $\lambda$ is called the Lagrangian multiplier. Therefore we get the following Lagrangian function
\begin{align*}
L(B[i], \Lambda) = \|D^TB[i]\|_F^2+\Lambda : \left(D[i]^TB[i]-I_d\right)
\end{align*}
with Lagrangian multiplier $\Lambda\in\R^{d\times d}$. Here $:$ is the Frobenius inner product defined as 
\begin{align*}
A:B = \sum_{i=1}^d\sum_{i=1}^d a_{i,j}b_{i,j}
\end{align*} 
for $A,B\in\R^{d\times d}$. Thus we get the necessary conditions
\begin{align*}
\nabla_{B[i]}L(B[i], \Lambda)&= 2D\left(D^TB[i]\right) + D[i]\Lambda = 0 \\
\nabla_{\Lambda}L(B[i], \Lambda)& = D[i]^TB[i]-I_d = 0
\end{align*}
which leads to the following equation system
\begin{align*}
       \left( \begin{array}{cc}
          2{D}{D}^T   &  D[i]\\
          D[i]^T  & 0
        \end{array} \right)
        \left( \begin{array}{c}
          	B[i]  \\
             \Lambda
        \end{array}
        \right) = \left(
             \begin{array}{c}
                  0  \\
                  I_d 
             \end{array}
        \right) \, .
    \end{align*}
    This system has the following minimum norm solution
\begin{align}
\left(\begin{array}{c}
B[i]\\ \Lambda 
\end{array}\right) =  \left( \begin{array}{cc}
          2{D}{D}^T   &  D[i]\\
          D[i]^T  & 0
        \end{array} \right)^+\left(\begin{array}{c}
        0 \\ I_d
        \end{array}\right)\, ,\label{eq::UP_solution}
\end{align} 
where $+$ denotes the Moore–Penrose inverse of a matrix \cite{ben2003generalized}. In \cite{hung1975moore} the Moore–Penrose inverse for a $2\times 2$ block partitioned matrix was derived.  Applying the first theorem in \cite{hung1975moore} to \eqref{eq::UP_solution} yields the statement.
\end{proof} 
\section{Proof of Theorem \ref{thm:mmvTrick1}}\label{proof:mmvTrick}
\begin{proof}
The proof is straight forward. Let $\tilde{B}$ attain the infimum in
\begin{align*}
    \underset{\underset{\underset{\forall 1\leq i \leq n}{\tilde{B}_{:,i}K_{:,i}=1}}{\tilde{B}\in\R^{m\times n}}}{inf}    \underset{i\neq j}{\text{max}}\, |\tilde{B}_{:,i}^TK_{:,j}|\, .
    \end{align*}
Let $B=\tilde{B}\otimes I_{d}$. We have
\begin{align*}
        (\tilde{B}\otimes I_{d})[i]^T(K\otimes I_{d})[i]& = 
        \left(\begin{array}{c}
            \tilde{b}^1_{1,i}I_{d}\\
            \hdots\\
            \tilde{b}^1_{m,i}I_{d}
        \end{array}\right)^T
        \left(\begin{array}{c}
            k_{1,i}I_{d}\\
            \hdots\\
            k_{m,i}I_{d}
        \end{array}\right)\\
        & = \left(\begin{array}{ccc}
            \tilde{b}^1_{1,i}I_d&
            \cdots&
            \tilde{b}^1_{m,i}I_d
        \end{array}\right)
        \left(\begin{array}{c}
            k_{1,i}I_{d}\\
            \hdots\\
            k_{m,i}I_{d}
        \end{array}\right)\\
        & = \tilde{B}^{T}_{:,i}K_{:,i}I_dI_{d} =I_{d}
\end{align*}
since $\tilde{B}^{T}_{:,i}K_{:,i}=1$, thus $B$ is a feasible solution. With the same argumentation we have
\begin{align*}
        \max_{i\neq j } \|B^T[i]D[j]\|_2
        & = \max_{i\neq j } \| 
        \left(\begin{array}{c}
             \tilde{b}_{1,i}I_d  \\
             \vdots\\
             \tilde{b}_{m,i}I_d
        \end{array}\right)^T
        \left(\begin{array}{c}
             k_{1,j}I_d  \\
             \vdots\\
             k_{m,j}I_d
        \end{array}\right)\|_2\\
        & = \max_{i\neq j } \| 
        \left(\begin{array}{ccc}
             \tilde{b}_{1,i}I_d  &
             \hdots &
             \tilde{b}_{m,i}I_d
        \end{array}\right)  
        \left(\begin{array}{c}
             k_{1,j}I_d  \\
             \vdots\\
             k_{m,j}I_d
        \end{array}\right)\|_2\\
        & = \max_{i\neq j } \|\left(\sum_{l=1}^{m}\tilde{b}_{l,i}k_{l,j}\right)I_d\|_2\\
        & = \max_{i\neq j } \|\left(\tilde{B}_{:,i}^TK_{:,j}\right)I_d\|_2\\
        & = \underset{i\neq j}{\text{max}}\, |\tilde{B}_{:,i}^TK_{:,j}|
    \end{align*}
taking the \textit{infimum} on both sides yields the result, without normalization of $\frac{1}{d}$.
\end{proof}
\section{Proof of Lemma 1}
\begin{proof}
We know that the solution of \eqref{reducedmutblockcoh} is given as the solution of the following equations (also in the sparse case)
\begin{align}
     \left( \begin{array}{cc}
          2{K}{K}^T   &  K_{:,i}\\
          K_{:,i}^T  & 0
        \end{array} \right)
        \left( \begin{array}{c}
          	B_{:,i}  \\
             \lambda
        \end{array}
        \right) = \left(
             \begin{array}{c}
                  0  \\
                  1
             \end{array}
        \right) \, .\label{eq::SolOrigUpperBound}
\end{align}
We know that $K$ is a circular matrix, this means the columns of $K$ are given as $K_{:,i} = T^{i-1} k$ where
\begin{align}
    T = \left(\begin{array}{cccc}
        0 &  \dots & 0 & 1  \\
        1 &  \dots & 0 & 0 \\
        \vdots & \ddots & \vdots & \vdots \\
        0 & \dots & 1 & 0 
    \end{array}\right)\, .
\end{align}
We now show that $B=circ(b)$, where $b$ solves \eqref{eq::SolUpperBoundCirc}, satisfies the linear system in \eqref{eq::SolOrigUpperBound}. 
\begin{align*}
    T^{-(i-1)}\left(2KK^TB_{:,i}+\lambda K_{:,i}\right) &= 2T^{-(i-1)}KK^TT^{(i-1)}b+\lambda T^{-(i-1)}T^{(i-1)}k\\
    & = 2KK^Tb+\lambda k = 0
\end{align*}
and therefore we have
\begin{align*}
    &T^{-(i-1)}\left(2KK^TB_{:,i}+\lambda K_{:,i}\right)=0\\
    \Rightarrow &2KK^TB_{:,i}+\lambda T^{-(i-1)}K_{:,i}=0
\end{align*}
{To finish the proof we have to solve the following problem: Since we solve this system for every $i$, the Lagrangian variable could change for every $i$.}
On the other hand we observe the following
\begin{align*}
    \lambda_i K_{:,i} &= -2 K K^T B_{:,i}  \\
    \Leftrightarrow \lambda_i K_{:,i}^TK_{:,i}&= -2K_{:,i}^T K K^T B_{:,i}\\
    \Leftrightarrow \lambda_i &= -2K_{:,i}^T K K^T B_{:,i}\\
    & = -2 (T^{i-1}k)^T K K^T T^{i-1}b\\
    & = -2 k^T T^{-(i-1)} K K^T T^{i-1}b\\
    & = -2 k^T K K^T b
\end{align*}
i.e. the lagrangian multiplier for $B=circ(b)$ does not change for $i$.\\
\end{proof}
\end{appendices}
\end{document}